\RequirePackage[T1]{fontenc}
\documentclass[12pt]{article}

\usepackage[height=8.85in,width=6.45in]{geometry}

\usepackage[utf8]{inputenc}
\usepackage{amsmath}
\usepackage{amssymb}
\usepackage{mathtools}
\numberwithin{equation}{section}
\usepackage{slashed}
\usepackage{braket}
\usepackage{enumitem}
\usepackage[svgnames]{xcolor}
\usepackage[colorlinks,citecolor=DarkGreen,linkcolor=FireBrick,urlcolor=FireBrick,linktocpage,unicode,psdextra]{hyperref}
\usepackage{cite}
\usepackage{graphicx}
\usepackage[bottom]{footmisc}

\usepackage{times}
\usepackage{courier}
\usepackage{bm}
\usepackage{subfig}
\usepackage{adjustbox}

\usepackage{colortbl}
\usepackage{xcolor}
\usepackage{mdframed}

\renewenvironment{figure}[1][]{
  \begin{originalfigure}[#1]
    \begin{mdframed}[linecolor=black!0,backgroundcolor=white]
}{
    \end{mdframed}
  \end{originalfigure}
}

\def\cA{\mathcal{A}}

\usepackage{amsthm}
\theoremstyle{plain}

\numberwithin{them}{section}

\theoremstyle{remark}

\def\ZZ{\mathbb{Z}}
\def\CC{\mathbb{C}}

\newcommand{\bea}{\begin{eqnarray}}
\newcommand{\eea}{\end{eqnarray}}
\def\no{\nonumber}
\def\m{\mu}

\def\cB{\mathcal{B}}

\def\cP{\mathcal{P}}
\def\cC{\mathcal{C}}
\def\cF{\mathcal{F}}
\def\cM{\mathcal{M}}
\def\cZ{\mathcal{Z}}
\def\cA{\mathcal{A}}
\def\cB{\mathcal{B}}
\def\cW{\mathcal{W}}
\def\cV{\mathcal{V}}
\def\cX{\mathcal{X}}

\def\eps{\epsilon}
\def\half{{1\over 2}}
\def\cN{\mathcal{N}}
\usepackage{dsfont}
\def\S{{\mathsf S}}
\def\T{{\mathsf T}}
\def\F{\mathsf{F}}
\def\FL{\mathsf{F_L}}

\usepackage{tikz}
\usetikzlibrary{decorations.markings}

\begin{document}

\begin{titlepage}

\begin{flushright}
KYUSHU-HET-360
\end{flushright}

\vskip 3cm

\begin{center}

{\Large \bfseries Parafermionizing the Monster}

\vskip 2cm

Yamato Honda$^1$,
Justin Kaidi$^{1,2,3}$,
Ippo Orii$^4$, 
\vskip 1cm

\begin{tabular}{ll}

1 & Department of Physics, Kyushu University, Fukuoka 819-0395, Japan \\
2& Institute for Advanced Study, Kyushu University, Fukuoka 819-0395, Japan\\
3 & Quantum and Spacetime Research Institute (QuaSR), Kyushu University, Fukuoka 819-0395, Japan
\\
4 & Kavli Institute for the Physics and Mathematics of the Universe (WPI), \\
& University of Tokyo, Kashiwa, Chiba 277-8583, Japan \\

\end{tabular}

\vskip 2cm

\end{center}

\noindent

We study the parafermionization of the Monster CFT with respect to its $\ZZ_{pA}$ subgroups, with $p$ an odd prime. 
Under certain assumptions, we show that the parafermionization is equal to a non-invertible gauging of $\cP(p) \times \cP(p)^\vee$, where $\cP(p)$ is the theory of $\ZZ_p$-parafermions and $\cP(p)^\vee$ is an appropriate dual theory, with global symmetry characterized by the centralizer of $\ZZ_{pA}$. By tracking the symmetries of $\cP(p) \times \cP(p)^\vee$ through the non-invertible gauging, we argue that the diagonal Monster CFT has $\mathrm{Rep}(\mathfrak{so}(3)_p) \boxtimes \mathrm{Rep}(\mathfrak{so}(3)_p)^\mathrm{op}$ symmetry, and hence that the holomorphic Monster theory has symmetry $\mathrm{Rep}(\mathfrak{so}(3)_p)$. We then compute the defect McKay-Thompson series associated to these symmetries, and prove that their invariance subgroups are $\Gamma_1(p+2)$.

\end{titlepage}

\setcounter{tocdepth}{2}
\tableofcontents

\section{Introduction}
\label{sec:introduction}

The Monster theory $V^\natural$ is a holomorphic conformal field theory of central charge $c=24$ whose automorphism group is the Monster group $\mathbb M$. Its character is the normalized modular invariant
\bea
J(\tau)=q^{-1}+196884\,q+21493760\,q^2+864299970\,q^3+\cdots~,
\eea
and the graded traces of Monster elements give the McKay-Thompson series of monstrous moonshine \cite{conway1979monstrous,Frenkel:1988xz,Borcherds:1992jjg}.  From the point of view of two-dimensional conformal field theory, a McKay-Thompson series is the torus partition function with a topological symmetry line inserted along the spatial cycle. Thus the Monster theory, and the associated diagonal conformal field theory $\cM:=V^\natural \otimes \overline{V^\natural}$,  provide a natural physical arena in which to understand the  phenomenon of moonshine. 

While the Monster theory is a bosonic CFT, one can also consider closely related spin-CFT generalizations.  Indeed, the Monster has two order-two conjugacy classes $2A$ and $2B$, with associated subgroups $\ZZ_{2A}$ and $\ZZ_{2B}$, and one can consider the fermionization of $\cM$ with respect to them.  For example, with respect to $\ZZ_{2A}$, one has the following simple relationship  \cite{Lin:2019hks}
\bea
\label{eq:firstrel}
\mathrm{Ferm}_{\ZZ_{2A}}(V^\natural) = \mathrm{MW}\times \mathrm{F}V\mathrm{BM}^\natural~,
\eea
or equivalently 
\bea
\label{eq:secondrel}
V^\natural = \mathrm{Bos}( \mathrm{MW}\times \mathrm{F}V\mathrm{BM}^\natural)~,
\eea
where $\mathrm{MW}$ is the theory of a single Majorana-Weyl fermion, $\mathrm{F}V\mathrm{BM}^\natural$ is the simple current extention of a certain theory $V\mathrm{BM}^\natural$  with Baby Monster symmetry $\mathds{BM}$ \cite{Hoehn:2007vut}, and both theories are holomorphic spin-CFTs.

However, as will be clarified below, at the level of diagonal conformal field theories, the second relationship is modified as,
\bea
\label{eq:Monsterfermionization}
\boxed{\,\,\cM= \mathrm{Bos}\left({\cM\cW} \times {\cF\cB\cM}\right)/\ZZ_2~,\,\,}
\eea
where $\cM \cW$ (resp. ${\cF\cB\cM}$) is the diagonal spin-CFT constructed from $\mathrm{MW}$ (resp. $\mathrm{F}V\mathrm{BM}^\natural$), and the $\ZZ_2 $ symmetry being gauged is defined in Section \ref{sec:Z2case}. The analog of the first relationship is then obtained by fermionizing both sides, giving 
\bea
\label{eq:Monsterfermionization2}
\boxed{\,\,\mathrm{Ferm}_{\ZZ_{2A}}\cM= ({\cM\cW} \times {\cF\cB\cM})/(-1)^{\mathsf{F}_L}_\mathrm{diag}~,\,\,}
\eea
where $(-1)^{\mathsf{F}_L}_\mathrm{diag}$ is diagonal left-moving fermion number.

The purpose of this note is to study the analogous story for parafermionization with respect to the subgroups $\ZZ_{pA}\subset \mathbb{M}$, where $p$ is an odd prime. 
The basic building block is the $\ZZ_p$-parafermion theory \cite{Fateev:1985mm}. Unlike $\mathrm{MW}$ or $\mathrm{F}V\mathrm{BM^\natural}$, which could be understood as holomorphic spin-CFTs, the natural $\ZZ_p$-analogs will be seen to be non-holomorphic paraspin-CFTs. Thus, instead of generalizing the holomorphic relations (\ref{eq:secondrel}) and  (\ref{eq:firstrel}), our goal will  be  to generalize the non-holomorphic relations  (\ref{eq:Monsterfermionization}) and (\ref{eq:Monsterfermionization2}).\footnote{Strictly speaking, throughout this paper our parafermionization relations are established at the level of torus partition functions with specified paraspin sectors, and we do not claim identities at the level of full paraspin-CFTs, if such a thing can even be defined. See Appendix \ref{app:fermionization} for a discussion of such subtleties. 
}

As such, we work with the diagonal $\ZZ_p$-parafermion theory $\cP(p)$, whose bosonization is the coset theory $\cB(p) = \mathfrak{su}(2)_p/\mathfrak{u}(1)$.  The primaries of $\cB(p)$ can be written as
\bea
(2\alpha,2\mu)~,\qquad \alpha=0,\dots,{p-1\over 2}~,\qquad \mu\in \ZZ_p~,
\eea
where for each fixed $\alpha$, the index $\mu$ labels a $\ZZ_p$ simple-current orbit.  The key assumption of our construction is the existence of a second $\ZZ_p$-equivariant rational theory, which we denote schematically by $\cB(p)^\vee$, with the same orbit structure,  and whose characters satisfy a bilinear relation of the form
\bea
\label{eq:bilinearassumption}
J(\tau) = \sum_{\alpha=0}^{{ (p-1)/ 2}} \boldsymbol\chi^{\cB(p)}_\alpha(\tau)\cdot \boldsymbol\chi^{\cB(p)^\vee}_\alpha(\tau)~,
\eea
where
\bea
\boldsymbol\chi^{\cB(p)}_\alpha :=  \big(\chi^{\cB(p)}_{(2\alpha,0)},\,\chi^{\cB(p)}_{(2\alpha,2)},\ldots, \,\chi^{\cB(p)}_{(2\alpha,2p-2)}\big)~,
\eea
and similarly for $\cB(p)^\vee$.

For $p=3,5,7$, the existence of such a decomposition is well-supported. In particular, for $p=3$ the role of $\cB(3)^\vee$ is played by the Fischer theory $\cB(3)^\vee = \cF_{24}$, known for its sporadic $3.\mathrm{Fi}'_{24}$ symmetry \cite{miyamoto20013,kitazume20033,hohn2012mckay,Bae:2018qfh,Bae:2020pvv}. For $p=5$, the theory $\cB(5)^\vee$ can be identified using the Harada-Norton decomposition of the Monster theory given in \cite{Bae:2020pvv}, as discussed in Appendix \ref{app:characters}.  For $p=7$, we may use the Held-group character data obtained from the Hecke image construction of \cite{Bae:2020pvv}, absorbing the nontrivial pairing matrix into the definition of the dual character vector; this is again discussed in Appendix \ref{app:characters}. For more general $p$, however,  the existence of $\cB(p)^\vee$ as a fully constructed rational CFT remains an assumption.\footnote{One reason to expect such a decomposition is that the Monster theory is known to be self-dual under gauging $\ZZ_{pA}$ \cite{Tuite:1993hy,Carnahan:2017pnh,abe2017remark}, and hence must contain a $\mathrm{TY}(\ZZ_p)$ duality symmetry \cite{Lin:2019hks,Volpato:2024goy}. It is natural to suspect that this comes from the known $\mathrm{TY}(\ZZ_p)$ symmetry of $\cB(p)$, in much the same way as the duality under gauging $\ZZ_{2A}$ comes from the Kramers-Wannier duality of the Ising theory.}

Assuming a decomposition of the above type, we will then show that the analog of (\ref{eq:Monsterfermionization}) for parafermionization of $\ZZ_{pA}$ is the following, 
\bea
\boxed{\,\,\cM =   \mathrm{Bos}(\cP(p) \times \cP(p)^\vee)/\cA^{(p)}_\mathrm{diag}~,\,\,}
\eea
where $\cP(p)^\vee$ is the parafermionization of $\cB(p)^\vee$.
Our main non-trivial result is the determination of the algebra $\cA^{(p)}_\mathrm{diag}$ appearing in this equality, which will involve non-invertible symmetries, and the use of this result to compute new defect McKay-Thompson series.

The basic idea is to reorganize the characters into the following combinations, 
\bea
\widetilde\chi_{\alpha,\beta}^{(p)} := \boldsymbol\chi^{\cB(p)}_\alpha\cdot
\boldsymbol\chi^{\cB(p)^\vee}_\beta ~,
\eea
in terms of which the assumed bilinear relation (\ref{eq:bilinearassumption}) can be written as
\bea 
|J(\tau)|^2 = \left|\, \sum_{\alpha=0}^{(p-1)/2} \widetilde\chi^{(p)}_{\alpha,\alpha}(\tau)\, \right|^2 ~. 
\eea
On the other hand, we will prove that the bosonization $\mathrm{Bos}(\mathcal P(p)\times \mathcal P(p)^\vee)$, which we denote by $\cB^{(p)}_\times$, is the diagonal theory
\bea
Z^{\cB^{(p)}_\times} = \sum_{\alpha,\beta=0}^{(p-1)/2} \left|\widetilde\chi^{(p)}_{\alpha,\beta}\right|^2~,
\eea
up to a harmless charge-conjugation refinement described below. Thus $\cM$ and  $\cB^{(p)}_\times$ are two different modular invariants constructed from the same underlying character vector.

The modular data of $\cB^{(p)}_\times$ is then shown to be that of the modular tensor category\footnote{Here and below $\boxtimes$ denotes the Deligne tensor product of fusion categories. The simple objects of  $\mathrm{Rep}(\mathfrak{so}(3)_p)\boxtimes \mathrm{Rep}(\mathfrak{so}(3)_p)^{op}$ are pairs $L_\alpha\boxtimes \overline{L_\beta}$, with componentwise fusion. The superscript $\mathrm{op}$ reverses the braiding, or equivalently inverts the $T$-matrix.}
\bea
\mathrm{Rep}(\mathfrak{so}(3)_p)\boxtimes \mathrm{Rep}(\mathfrak{so}(3)_p)^\mathrm{op}~,
\eea
i.e. the representation ring of integrable highest-weight representations of the affine Lie algebra $\mathfrak{su}(2)_p$, restricted to
integral spins, times its opposite. 
From this, we see that the Monster modular invariant can be obtained by gauging the canonical diagonal algebra
\bea
\cA^{(p)}_{\rm diag} = \bigoplus_{\alpha=0}^{(p-1)/2} L_\alpha\boxtimes \overline L_\alpha ~,
\eea
where gauging here is meant in the general sense, appropriate for non-invertible symmetries \cite{Frohlich:2009gb,Diatlyk:2023fwf,Perez-Lona:2023djo,Perez-Lona:2024sds}. 
Conversely, gauging the dual diagonal algebra in $\cM$ returns us to the $\cB^{(p)}_\times$ theory.  This allows us to conclude that the diagonal Monster CFT inherits a dual $\mathrm{Rep}(\mathfrak{so}(3)_p)\boxtimes \mathrm{Rep}(\mathfrak{so}(3)_p)^\mathrm{op}$ symmetry, and consequently that $V^\natural$ itself has $\mathrm{Rep}(\mathfrak{so}(3)_p)$ symmetry.

Having shown this, we then use this symmetry to define defect McKay-Thompson series
\bea
T^{(p)}_{L_\alpha}(\tau) = \mathrm{Tr}_{\mathcal{H}}\!\left(\hat L_\alpha\,q^{L_0-c/24}\right)~,
\eea
which we give explicitly for $p=3,5,7$.  The $p=3$ case reproduces the Fibonacci defect McKay-Thompson series found previously in \cite{Fosbinder-Elkins:2024hff,Moller:2024xtt}, while for $p=5$ and $p=7$ we obtain new defect McKay-Thompson series associated to the nontrivial simple objects of $\mathrm{Rep}(\mathfrak{so}(3)_5)$ and $\mathrm{Rep}(\mathfrak{so}(3)_7)$.

Finally, we determine the modular invariance groups of these defect McKay-Thompson series in $\mathrm{PSL}(2, \ZZ)$.  Defining
\bea
\Gamma^{(p)}_\alpha = \left\{\gamma\in \mathrm{PSL}(2,\mathbb Z)\;|\; T^{(p)}_{L_\alpha}(\gamma\tau)=T^{(p)}_{L_\alpha}(\tau)\right\}~, 
\eea
we prove that for every nontrivial line $L_\alpha$, we have
\bea
\Gamma^{(p)}_\alpha=\Gamma_1(p+2)~.
\eea
The inclusion $\Gamma_1(p+2)\subseteq \Gamma^{(p)}_\alpha$ is a relatively straightforward consequence of the congruence property of modular tensor categories \cite{Ng:2010win,Ng:2012ty}, while the reverse inclusion uses an explicit finite-Weil realization of the $\mathfrak{so}(3)_p$ modular representation.  
For the cases of $p=3,5,7$, the groups $\Gamma_1(p+2)$ are genus-zero subgroups, but our result makes clear that genus zero is a low-lying accident rather than a general consequence of the construction. Indeed, the groups $\Gamma_1(N)$ have genus zero only for $N\leq 10$ or $N=12$, which includes only the lowest-lying cases above. For larger primes, if the corresponding decomposition exists, the resulting invariance groups will generally have positive genus.  

\paragraph{Organization:}
This note is organized as follows.  In Section \ref{sec:warmup}, we recall the fermionization of the holomorphic Monster theory $V^\natural$ with respect to $2A$, extend the discussion to the diagonal Monster CFT $\cM$, and then generalize to parafermionization with respect to $3A$, explaining the appearance of non-invertible gauging. This serves as a warm-up for the  general $\ZZ_{pA}$ construction, which is given in Section \ref{sec:genericp}.  Then in Section \ref{sec:defectMTseries}, we compute the corresponding defect McKay--Thompson series and prove that their invariance groups are $\Gamma_1(p+2)$.  

We also include several appendices. Appendix \ref{app:fermionization} provides a brief review of fermionization and parafermionization. Appendix \ref{app:characters} collects the explicit character data for $p=3,5,7$, used to compute defect McKay-Thompson series. Finally, Appendix \ref{app:modular} derives various modular transformation formulas used in the proof of the invariance subgroup result.

\paragraph{Notation:} 
Throughout this note, we will use $V^\natural$, $V\mathrm{BM}^\natural$, $V\mathrm{F}_{24}^\natural$ to denote VOAs or holomorphic (spin-)CFTs, while the corresponding diagonal CFTs will be denoted as $\cM$, $\cB\cM, \cF_{24}$. 
Likewise, the notation $\cP(p)$, $\cB(p)$ will be reserved for diagonal $\ZZ_p$-parafermion CFTs and their bosonizations. For convenience, we label characters by their corresponding diagonal CFTs, giving $\chi^{\cM}(\tau), \chi^{\cB(p)}(\tau)$, and so on.

\section{Warm-up: Fermionizing $\ZZ_{2A}$ and parafermionizing $\ZZ_{3 A}$}
\label{sec:warmup}

\subsection{$\ZZ_{2A}$ fermionization}
\label{sec:Z2case}

We begin by recalling the fermionization of the Monster with respect to the $2A$ involution, beginning at the level of holomorphic theories \cite{Lin:2019hks}.  Let
\bea
g\in 2A\subset \mathbb{M}~, \hspace{0.2 in}  \ZZ_{2A}=\langle g\rangle ~
\eea
and denote the corresponding twisted-twined partition functions by
\bea
Z^{\cM}_{a,b}(\tau):=\mathrm{Tr}_{\mathcal H_{g^a}} \left(g^b q^{L_0-c/24}\right)~, \hspace{0.4in} a,b\in  \ZZ_2~,
\eea
so that we have
\bea
Z^{\cM}_{0,0}(\tau)=J(\tau)~
\eea
as well as
\bea
Z^{\cM}_{0,1}(\tau)=T_{2A}(\tau) = \left( \left({\eta(\tau)\over \eta(2\tau)}\right)^{12} + 2^6\left({\eta(2\tau) \over \eta(\tau)}\right)^{12} \right)^2 -104 ~. 
\eea
The remaining two sectors are obtained by modular transformations,
\bea
Z^{\cM}_{1,0}(\tau)=Z^{\cM}_{0,1}(-1/\tau)~, \hspace{0.4 in} Z^{\cM}_{1,1}(\tau)=Z^{\cM}_{1,0}(\tau+1)~,
\eea
giving explicitly,
\bea
Z^{\cM}_{1,0}(\tau)&=& \left( 2^6\left(\frac{\eta(\tau)}{\eta(\tau/2)}\right)^{12} + \left(\frac{\eta(\tau/2)}{\eta(\tau)}\right)^{12} \right)^2 -104~,
\no\\
Z^{\cM}_{1,1}(\tau)&=& \left( 2^6\left(\frac{\eta(\tau)}{\eta(\tau/2+1/2)}\right)^{12} + \left(\frac{\eta(\tau/2+1/2)}{\eta(\tau)}\right)^{12} \right)^2 -104~.
\eea

These results may be used to compute the fermionized partition function (see Appendix \ref{app:fermionization} for a brief review of fermionization),
\bea
\label{eq:monsterfermionization}
Z^{\cF}_{s_1,s_2} = \half \sum_{a,b\in\ZZ_2} (-1)^{(s_1+a)(s_2+b)} Z^{\cM}_{a,b}~;
\eea
for example, the NSNS partition function is given by
\bea
\label{eq:NSNSchar}
Z^{\cF}_\mathrm{NSNS}(\tau) &:=& Z^{\cF}_{0,0}(\tau) \,\,=\,\,{1\over 2} \left( J(\tau)+Z^{\cM}_{0,1}(\tau)+Z^{\cM}_{1,0}(\tau)-Z^{\cM}_{1,1}(\tau) \right)
\no\\
 &=& q^{-1}+q^{-1/2}+4372\,q^{1/2}+100628\,q+\cdots ~.
\eea
In \cite{Lin:2019hks}, this was interpreted as the product
\bea
\label{eq:LinShaoclaim}
\mathrm{Ferm}_{\ZZ_{2A}}(V^\natural) = \mathrm{MW}\times \mathrm{F}V\mathrm{BM}^\natural~,
\eea
where $\mathrm{MW}$ is a chiral Majorana-Weyl fermion and $\mathrm{F}V\mathrm{BM}^\natural$ is the fermionization of the Baby Monster VOA \cite{Hoehn:2007vut}; both of these are holomorphic spin-CFTs. 

One route towards this identification is the fact that the Monster character admits the following bilinear decomposition \cite{yamauchi2002z_2,Hampapura:2016mmz},
\bea
\label{eq:Z2bilinearrel}
J(\tau) = \chi^{\mathrm{Ising}}_0\,\chi^{\cB\cM}_0 + \chi^{\mathrm{Ising}}_{1/2}\,\chi^{\cB\cM}_{3/2} + \chi^{\mathrm{Ising}}_{1/16}\,\chi^{\cB\cM}_{31/16}~,
\eea
where  $\chi^{\mathrm{Ising}}_h$ and $\chi^{\cB\cM}_h$ are characters for the Ising and Baby Monster theories. 
The Ising portion fermionizes to a Majorana-Weyl fermion, whose spin-sector partition functions are
\bea
\label{eq:MWIsingrelation}
Z^{\cM\cW}_{\mathrm{NSNS}} = \chi^{\mathrm{Ising}}_0+\chi^{\mathrm{Ising}}_{1/2}~,  \hspace{0.2 in} Z^{\cM\cW}_{\mathrm{NSR}} = \chi^{\mathrm{Ising}}_0-\chi^{\mathrm{Ising}}_{1/2}~, \hspace{0.2 in} Z^{\cM\cW}_{\mathrm{RNS}} = \sqrt{2}\,\chi^{\mathrm{Ising}}_{1/16}~,  \hspace{0.2 in} Z^{\cM\cW}_{\mathrm{RR}}=0~, 
\no\\
\eea
with exactly analogous expressions holding for the fermionized Baby Monster. 
Therefore, the bilinear relation above can be rewritten as the bosonization of the product $\mathrm{MW}\times\mathrm{F}V\mathrm{BM}^\natural$, namely
\bea
J(\tau) = \half \left( Z^{\cM\cW}_{\mathrm{NSNS}}Z^{\cF\cB\cM}_{\mathrm{NSNS}} + Z^{\cM\cW}_{\mathrm{NSR}}Z^{\cF\cB\cM}_{\mathrm{NSR}} + Z^{\cM\cW}_{\mathrm{RNS}}Z^{\cF\cB\cM}_{\mathrm{RNS}} - Z^{\cM\cW}_{\mathrm{RR}}Z^{\cF\cB\cM}_{\mathrm{RR}} \right)~.
\eea
Conversely, the NSNS partition function of the fermionized Monster $\cF$ is simply
\bea
Z^{\cF}_{\mathrm{NSNS}} = Z^{\cM\cW}_{\mathrm{NSNS}}Z^{\cF\cB\cM}_{\mathrm{NSNS}} = (\chi^{\mathrm{Ising}}_0+\chi^{\mathrm{Ising}}_{1/2}) (\chi^{\cB\cM}_0+\chi^{\cB\cM}_{3/2})~,
\eea
which indeed reproduces the result in (\ref{eq:NSNSchar}).
Thus the result of \cite{Lin:2019hks} can be alternatively summarized as
\bea  
\label{eq:MonsterVOAstatement}
V^\natural = \mathrm{Bos}\left(\mathrm{MW}\times\mathrm{F}V\mathrm{BM}^\natural\right)~.
\eea

The above discussion has been entirely at the level of holomorphic theories. Next let us consider the fermionization of $\cM= V^\natural \otimes \overline{V^\natural}$, whose partition function is 
\bea
Z^\cM = |J(\tau)|^2~. 
\eea
On the one hand, the bilinear relation (\ref{eq:Z2bilinearrel}) tells us that we have 
\bea
Z^\cM = | \chi^{\mathrm{Ising}}_0\,\chi^{\cB\cM}_0 + \chi^{\mathrm{Ising}}_{1/2}\,\chi^{\cB\cM}_{3/2} + \chi^{\mathrm{Ising}}_{1/16}\,\chi^{\cB\cM}_{31/16}|^2~.
\eea
On the other hand, if we denote the diagonal spin-CFTs constructed from $\mathrm{MW}$ and $\mathrm{F}V\mathrm{BM}^\natural$ as $\cM \cW$ and $\cF\cB \cM$, respectively, and consider the bosonization $\cB_\times^{(2)} := \mathrm{Bos}( \cM \cW \times \cF\cB \cM)$, we obtain 
\bea
Z^{\cB_\times^{(2)}} &=& \half \left( |Z^{\cM \cW}_{\mathrm{NSNS}}Z^{\cF\cB \cM}_{\mathrm{NSNS}}| ^2 + |Z^{\cM \cW}_{\mathrm{NSR}}Z^{\cF\cB \cM}_{\mathrm{NSR}}| ^2+ |Z^{\cM \cW}_{\mathrm{RNS}}Z^{\cF\cB \cM}_{\mathrm{RNS}}| ^2+ |Z^{\cM \cW}_{\mathrm{RR}}Z^{\cF\cB \cM}_{\mathrm{RR}}| ^2 \right) 
\no\\
&=& |\chi^{\mathrm{Ising}}_{0} \chi^{\cB\cM}_{0} +\chi^{\mathrm{Ising}}_{1/2} \chi^{\cB\cM}_{3/2}  |^2 + |\chi^{\mathrm{Ising}}_{0} \chi^{\cB\cM}_{3/2} +\chi^{\mathrm{Ising}}_{1/2} \chi^{\cB\cM}_{0}|^2 + 2 |\chi^{\mathrm{Ising}}_{1/16}\,\chi^{\cB\cM}_{31/16}|^2 ~.\hspace{0.2 in}
\eea
Thus, the naive uplift of the relation (\ref{eq:MonsterVOAstatement}) to diagonal CFTs, namely 
\bea
\cM \stackrel{?}{=} \mathrm{Bos}\left(\cM\cW\times\cF \cB \cM\right)~,
\eea
 does not hold.

To understand the correct relation, consider the following three theories, 
\bea
\cM~, \hspace{0.2 in} \cB_\times^{(2)} = \mathrm{Bos}( \cM \cW \times \cF\cB \cM)~, \hspace{0.2 in} \cX = \mathrm{Ising} \times \cB \cM~,
\eea  
where $\cX$ is the diagonal nine-character bosonic CFT with characters 
\bea
\chi^{\mathrm{Ising}}_{\alpha} \chi^{\cB\cM}_{\beta}~. 
\eea 
In fact, since $\cM$ and $\cB_\times^{(2)}$ can also be written in terms of products $\chi^{\mathrm{Ising}}_{\alpha} \chi^{\cB\cM}_{\beta}$, it is clear that all three of these theories are just distinct modular invariants built from a single underlying vector of characters. These modular invariants can all be related by gauging different algebra objects in $\cX$.

The relevant gaugings are easily identified by recalling that the theory $\cX$ has symmetry $\mathrm{Ising} \boxtimes \mathrm{Ising}^\mathrm{op}$, coming from the symmetries of $\mathrm{Ising}$ and $\cB\cM$. This category has two non-trivial bosonic condensable algebras \cite{Chen:2019zcu,Bhardwaj:2024ydc}, 
\bea
\label{eq:twoalgebraobjects}
\cA_1 := 1 \boxtimes \bar 1 \oplus  \eta \boxtimes \bar \eta \oplus \cN \boxtimes \bar \cN~, \hspace{0.3 in} \cA_2:= 1 \boxtimes \bar 1 \oplus  \eta \boxtimes \bar \eta~, 
\eea
where the former is Lagrangian and the latter is not. We thus conclude that 
\bea
\label{eq:A1A2eqs}
\cM  = \cX / \cA_1 \hspace{0.3 in} \mathrm{and} \hspace{0.3 in} \cB_\times^{(2)}  = \cX / \cA_2~.
\eea
Furthermore, upon gauging $\cA_2$, the resulting module category is the toric code \cite{Chen:2019zcu}, and gauging either of the Lagrangian algebras $\cA_{e,m} = 1\oplus e$ or $1 \oplus m$ in that theory  (both of which are invertible) will again give $\cM$, i.e.
\bea
\cM  =\cB_\times^{(2)}  / \cA_{e,m} ~. 
\eea
Thus the correct way to lift (\ref{eq:MonsterVOAstatement}) to a statement about diagonal CFTs is the following,\bea
\label{eq:MonsterZ2mainrel}
\boxed{\,\, \cM = \mathrm{Bos}( \cM \cW \times \cF\cB \cM) /  \cA_{e,m}~.\,\,}
\eea

There is also a useful fermionic interpretation of this result.  Let $(-1)^{\F}_{\cM\cW}$ and $(-1)^{\FL}_{\cM\cW}$ be the total and left-moving fermion parities of $\cM\cW$, and likewise for $\cF\cB\cM$. We begin by bosonizing each piece separately to obtain the theory $\cX$, which has symmetry $\mathrm{Ising} \boxtimes \mathrm{Ising}^\mathrm{op}$. Under such bosonization, the symmetries map as follows \cite{Thorngren:2018bhj},
\bea
(-1)^{\F}_{\cM\cW} &\mapsto& \eta \boxtimes \bar 1~, \hspace{0.35 in}(-1)^{\FL}_{\cM\cW} \,\,\mapsto\,\, \cN \boxtimes \bar 1~, 
\no\\
(-1)^{\F}_{\cF\cB\cM} &\mapsto& 1 \boxtimes \bar \eta~, \hspace{0.3 in}(-1)^{\FL}_{\cF\cB\cM} \,\,\mapsto\,\, 1 \boxtimes \overline \cN~.
\eea
Next we gauge the algebra object $ \cA_2$ in (\ref{eq:twoalgebraobjects}) to go to $\cB_\times^{(2)}$. Under $\cA_2$, the objects $\eta \boxtimes 1$ and $1 \boxtimes \bar \eta$ are exchanged, and hence become a single object in $\cB_\times^{(2)}$, giving the toric code fermion $f$. On the other hand, the object $\cN \boxtimes \overline \cN$ is fixed under $\cA_2$, and hence upon gauging it is expected to split into two defects, which give the $e$ and $m$ lines of $\cB_\times^{(2)}$. In other words, we have 
\bea
(-1)^{\FL}_\mathrm{diag} \,\,\stackrel{\mathrm{bosonize}}{\longrightarrow}\,\, \cN \boxtimes \overline \cN \,\,\stackrel{\mathrm{gauge}\,\cA_2}{\longrightarrow}\,\, e \oplus m~,
\eea
where $(-1)^{\FL}_\mathrm{diag}:=(-1)^{\FL}_{\cM\cW}(-1)^{\FL}_{\cF\cB\cM}$. This implies that, upon fermionizing both sides of (\ref{eq:MonsterZ2mainrel}), we obtain the alternative expression
 \bea
 \label{eq:alternativeMonstermain}
 \boxed{\,\, \mathrm{Ferm}_{\ZZ_{2A}}\cM = ( \cM \cW \times \cF\cB \cM) / (-1)^{\FL}_\mathrm{diag}~.\,\,}
 \eea

This example provides the template for the parafermionic constructions below.  Starting with the $\ZZ_{3A}$ case, the role of the Ising model will be played by the three-state Potts model $\cB(3)$, whose parafermionization is the theory $\cP(3)$ of $\ZZ_3$-parafermions, while the role of the Baby Monster theory $\cB \cM$ will be played by the Fischer theory $\cB(3)^\vee = \cF_{24}$, whose parafermionization is denoted by $\cP(3)^\vee$.

\subsection{Interlude: fermionization vs. parafermionization}
\label{sec:clarification}

Before moving on, let us make some brief comments on the distinction between fermionization and parafermionization (additional comments can be found in Appendix \ref{app:fermionization}). Above, we have encountered two somewhat different notions of fermionization, namely: \textit{i)} fermionization of a bosonic CFT to obtain a spin-CFT, acting at the level of partition functions as in (\ref{eq:monsterfermionization}) or (\ref{eq:fermionizationdef}), and \textit{ii)} fermionization of a VOA, which is an order-2 simple current extension to obtain an sVOA. 

The two concepts are distinct, but closely related. For example, consider the extension of the Ising VOA to the  $\mathrm{MW}$ sVOA. The diagonal spin-CFT $\cM \cW$ constructed from $\mathrm{MW}$ has the following partition functions, 
\bea
\label{eq:diagonalMW}
Z_\mathrm{NSNS}^{\cM\cW} = |\chi^{\mathrm{Ising}}_0+\chi^{\mathrm{Ising}}_{1/2}|^2~, \hspace{0.2 in}Z_\mathrm{NSR}^{\cM\cW} = |\chi^{\mathrm{Ising}}_0-\chi^{\mathrm{Ising}}_{1/2}|^2 , \hspace{0.2 in} Z^{\cM\cW}_{\mathrm{RNS}} =2|\chi^{\mathrm{Ising}}_{1/16}|^2 ~,  \hspace{0.2 in} Z^{\cM\cW}_{\mathrm{RR}}=0~,
\no\\
\eea
as follows from the relations in (\ref{eq:MWIsingrelation}). 
On the other hand, starting from the Ising VOA, we may construct the diagonal Ising CFT, with twisted-twined partition functions, 
\bea
Z^\mathrm{Ising}_{0,0} &=& |\chi^{\mathrm{Ising}}_0|^2 + |\chi^{\mathrm{Ising}}_{1/2}|^2 + |\chi^{\mathrm{Ising}}_{1/16}|^2~,
\no\\
Z^\mathrm{Ising}_{0,1} &=& |\chi^{\mathrm{Ising}}_0|^2 + |\chi^{\mathrm{Ising}}_{1/2}|^2 - |\chi^{\mathrm{Ising}}_{1/16}|^2~,
\no\\
Z^\mathrm{Ising}_{1,0} &=& \chi^{\mathrm{Ising}}_0 \bar \chi^{\mathrm{Ising}}_{1/2} + \chi^{\mathrm{Ising}}_{1/2} \bar \chi^{\mathrm{Ising}}_{0}+ |\chi^{\mathrm{Ising}}_{1/16}|^2~,
\no\\
Z^\mathrm{Ising}_{1,1} &=& -\chi^{\mathrm{Ising}}_0 \bar \chi^{\mathrm{Ising}}_{1/2} - \chi^{\mathrm{Ising}}_{1/2} \bar \chi^{\mathrm{Ising}}_{0}+ |\chi^{\mathrm{Ising}}_{1/16}|^2~.
\eea
Then fermionizing this diagonal Ising CFT using (\ref{eq:fermionizationdef}), we obtain a set of partition functions which coincide precisely with (\ref{eq:diagonalMW}). Thus the fermionization (in the first sense) of the diagonal Ising CFT coincides with the diagonal spin-CFT constructed from the fermionization (in the second sense) of the Ising VOA. 
More generally, consider a VOA $V$ which admits an order-2 simple current extension to an sVOA $FV$. Then considering the corresponding diagonal CFT $\cV$ and spin-CFT $\cF \cV$, one may show that $\cF \cV$ is obtained from $\cV$ via fermionization in the first sense; see Appendix \ref{app:newappendix}. 

An important ingredient in the previous subsection was that the sVOAs $\mathrm{MW}$ and $\mathrm{F}V\mathrm{BM^\natural}$ gave rise to holomorphic spin-CFTs. Indeed, while a holomorphic bosonic CFT has a single modular invariant character,  a holomorphic spin-CFT has a single character in each spin sector,  transforming by the modular action,
\bea
&\S:& \left( \begin{matrix} Z^{\cF \cV}_\mathrm{NSNS} \\ Z^{\cF \cV}_\mathrm{NSR} \\ Z^{\cF \cV}_\mathrm{RNS}   \end{matrix}  \right) \mapsto \left(\begin{matrix} 1 & 0 & 0 \\ 0 & 0 & 1 \\ 0 & 1 & 0 \end{matrix}  \right)  \left( \begin{matrix} Z^{\cF \cV}_\mathrm{NSNS} \\ Z^{\cF \cV}_\mathrm{NSR} \\ Z^{\cF \cV}_\mathrm{RNS}   \end{matrix}  \right) ~, 
\no\\
&\T: & \left( \begin{matrix} Z^{\cF \cV}_\mathrm{NSNS} \\ Z^{\cF \cV}_\mathrm{NSR} \\ Z^{\cF \cV}_\mathrm{RNS}   \end{matrix}  \right) \mapsto e^{- 2 \pi i c_L / 24}\left(\begin{matrix} 0 & 1 & 0 \\ 1 & 0 & 0 \\ 0 & 0 & e^{2\pi i c_L /8} \end{matrix}  \right)  \left( \begin{matrix} Z^{\cF \cV}_\mathrm{NSNS} \\ Z^{\cF \cV}_\mathrm{NSR} \\ Z^{\cF \cV}_\mathrm{RNS}   \end{matrix}  \right) ~,  
\eea
and this is satisfied for (\ref{eq:MWIsingrelation}).
It is this holomorphic nature that allows for relations such as (\ref{eq:LinShaoclaim})  and (\ref{eq:MonsterVOAstatement}) between holomorphic (spin-)CFTs. 

Let us contrast the above with the case of parafermionization. In \cite{Yao:2020dqx,Thorngren:2021yso}, parafermionization was defined  at the level of torus partition functions, with several potential obstructions to its extension to higher genus; some of these obstructions are reviewed in Appendix \ref{app:fermionization}. As such, in this work we focus exclusively on torus partition functions. In addition, in contrast to fermionization, it is easy to see that if we consider a diagonal CFT $\cV$ coming from a VOA $V$, then the parafermionization $\cP \cV$ of $\cV$ \textit{cannot} always be interpreted as the diagonal paraspin-CFT constructed from a tentative $\ZZ_p$-graded chiral extension $PV$; see Appendix \ref{app:newappendix}. 

Crucial for our analysis below will be the fact that the obvious $\ZZ_p$ analogs to $\mathrm{MW}$ and $\mathrm{F}V\mathrm{BM}^\natural$ in fact do {not} define holomorphic paraspin-CFTs. Indeed,  in addition to the modular action on the paraspin sectors, we will see that there is modular data corresponding to $\mathrm{Rep}(\mathfrak{so}(3)_p)$, which means that there is more than one primary in each paraspin sector. As such, instead of trying to generalize the holomorphic relations in (\ref{eq:LinShaoclaim})  and (\ref{eq:MonsterVOAstatement}), our goal will be to generalize the non-holomorphic relations in (\ref{eq:MonsterZ2mainrel}) and (\ref{eq:alternativeMonstermain}).

\subsection{$\ZZ_{3A}$ parafermionization}
\label{sec:Z3warmup}

We now consider the {parafermionization} of the Monster theory with respect to $\ZZ_{3A}$. 
Our starting point is again a bilinear decomposition \cite{miyamoto20013,kitazume20033,hohn2012mckay,Bae:2018qfh,Bae:2020pvv},
\bea
\label{eq:bilinear2}
J(\tau) = \chi_0^\mathrm{3SP} \chi_0^{\cF_{24}} + \chi_{2 \over 5}^\mathrm{3SP} \chi_{8 \over 5}^{\cF_{24}} + 2\, \chi_{2\over 3}^\mathrm{3SP} \chi_{4\over 3}^{\cF_{24}} + 2\, \chi_{1\over 15}^\mathrm{3SP} \chi_{29\over 15}^{\cF_{24}}~,
\eea
where $\chi_h^\mathrm{3SP}$ are characters of the three-state Potts model for primaries of conformal weight $h$, while $\chi_h^{\cF_{24}}$ are a set of characters obtained from these by application of an appropriate Hecke operator---details are given in Appendix \ref{app:p3}. These characters are expected to correspond to a legitimate VOA $VF_{24}^\natural$  (and consequently a legitimate diagonal CFT $\cF_{24}$) with global  symmetry given by $3.\mathrm{Fi}_{24}'$, with $\mathrm{Fi}_{24}'$ an index 2 subgroup of the largest Fischer group $\mathrm{Fi}_{24}$ \cite{Bae:2020pvv}.

This suggests, in analogy to the case of fermionization described above, that parafermionization of the Monster theory can be built from the parafermionization of the three-state Potts model---namely the theory $\cP(3)$ of $\ZZ_3$-parafermions---times the parafermionization of $\cF_{24}$, which we denote by $\cP(3)^\vee$.
To this effect, note that the partition functions $Z^{\cP(3)}_{s_1,s_2}$ of $\cP(3)$ for various paraspin structures $(s_1, s_2)$, $s_i \in \ZZ_3$, can be computed using (\ref{eq:parafermionization}) in Appendix \ref{app:fermionization}, with the following results \cite{Yao:2020dqx},
\bea
Z^{\cP(3)}_{0,0} &=& | \chi_0^\mathrm{3SP}|^2 + | \chi_{2/5}^\mathrm{3SP}|^2 +2 \chi_0^\mathrm{3SP} \overline  \chi_{2/3}^\mathrm{3SP} + 2 \chi_{2/5}^\mathrm{3SP} \overline  \chi_{1/15}^\mathrm{3SP}~,
\no\\
Z^{\cP(3)}_{0,1} = Z^{\cP(3)}_{0,2} &=& | \chi_0^\mathrm{3SP}|^2 + | \chi_{2/5}^\mathrm{3SP}|^2 - \chi_0^\mathrm{3SP} \overline  \chi_{2/3}^\mathrm{3SP} -  \chi_{2/5}^\mathrm{3SP} \overline  \chi_{1/15}^\mathrm{3SP}~,
\no\\
Z^{\cP(3)}_{1,0} = Z^{\cP(3)}_{2,0} &=& 2 |\chi_{2/3}^\mathrm{3SP}|^2 +2 |\chi_{1/15}^\mathrm{3SP}|^2 +\chi_{2/3}^\mathrm{3SP} \overline \chi_{0}^\mathrm{3SP} + \chi_{1/15}^\mathrm{3SP} \overline \chi_{2/5}^\mathrm{3SP}~,
\no\\
Z^{\cP(3)}_{1,1} = Z^{\cP(3)}_{2,2} &=&(1+e^{2\pi i \over 3})( |\chi_{2/3}^\mathrm{3SP}|^2 +|\chi_{1/15}^\mathrm{3SP}|^2) + e^{4 \pi i \over 3}(\chi_{2/3}^\mathrm{3SP} \overline \chi_{0}^\mathrm{3SP} + \chi_{1/15}^\mathrm{3SP} \overline \chi_{2/5}^\mathrm{3SP})~,
\no\\
Z^{\cP(3)}_{1,2} = Z^{\cP(3)}_{2,1} &=&(1+e^{4\pi i \over 3})( |\chi_{2/3}^\mathrm{3SP}|^2 +|\chi_{1/15}^\mathrm{3SP}|^2) + e^{2 \pi i \over 3}(\chi_{2/3}^\mathrm{3SP} \overline \chi_{0}^\mathrm{3SP} + \chi_{1/15}^\mathrm{3SP} \overline \chi_{2/5}^\mathrm{3SP})~.
\eea
Exactly analogous expressions hold for the parafermionization $\cP(3)^\vee$ of the Fischer CFT. 
If we now consider the bosonization  $\cB_\times^{(3)} := \mathrm{Bos}(\cP(3)\times \cP(3)^\vee)$, it has partition function
\bea
Z^{\cB_\times^{(3)}}(\tau)&=&{1\over 3}\sum_{s_1,s_2\in \ZZ_3}  e^{- {2 \pi i \over 3} s_1 s_2} \, Z^{\cP(3)}_{s_1,s_2}(\tau)\times Z^{\cP(3)^\vee}_{s_1,s_2}(\tau)
\no\\
&=& {1\over q \bar q}\left(1 + q^{2/5} \bar q^{2/5}+ 1568 \, q^{7/5} \bar q^{2/5} + 1568\, q^{2/5} \bar q^{7/5} + 2458624\, q^{7/5}\bar q^{7/5} +  \dots\right)~.\hspace{0.5 in}
\eea
The goal is to relate this theory to $\cM$ by appropriate discrete gauging.

To do so, note that if we define the following four characters 
\bea
\label{eq:Z3tildedchars}
\widetilde \chi^{(3)}_0 &:=& \chi_0^{\mathrm{3SP}} \chi_0^{\cF_{24}} +2 \chi_{2/3}^{\mathrm{3SP}} \chi_{4/3}^{\cF_{24}} ~,
\no\\
\widetilde \chi^{(3)}_{2/5} &:=& \chi_{2/5}^{\mathrm{3SP}} \chi_0^{\cF_{24}} +2 \chi_{1/15}^{\mathrm{3SP}} \chi_{4/3}^{\cF_{24}} ~,
\no\\
\widetilde \chi^{(3)}_{8/5} &:=& \chi_0^{\mathrm{3SP}} \chi_{8/5}^{\cF_{24}} +2 \chi_{2/3}^{\mathrm{3SP}} \chi_{29/15}^{\cF_{24}} ~,
\no\\
\widetilde \chi^{(3)}_{2} &:=& \chi_{2/5}^{\mathrm{3SP}} \chi_{8/5}^{\cF_{24}} +2 \chi_{1/15}^{\mathrm{3SP}} \chi_{29/15}^{\cF_{24}} ~,
\eea
then on the one hand we may write 
\bea
\label{eq:Z3chitildediagonal}
Z^{\cB_\times^{(3)}}= |\widetilde \chi^{(3)}_0|^2 + |\widetilde \chi^{(3)}_{2/5}|^2 + |\widetilde \chi^{(3)}_{8/5}|^2 + |\widetilde \chi^{(3)}_{2}|^2~,
\eea
while on the other hand we have
\bea
\label{eq:MonstertildechiZ3}
|J(\tau)|^2 = |\widetilde \chi^{(3)}_0 + \widetilde \chi^{(3)}_2|^2~. 
\eea
This means that the Monster CFT and $\cB_\times^{(3)}$ correspond to two separate modular invariants for a single four character vector. 

We may understand this in more detail by computing the modular $S$-matrix of $\cB_\times^{(3)}$, starting from the $S$-matrix $\S^{3\mathrm{SP}}$ of the three-state Potts model given in (\ref{eq:appSmat}) of Appendix \ref{app:characters}. In particular, the $S$-matrix of the tensor product theory $\mathrm{3SP} \times \cF_{24}$ is given by $\S^{3\mathrm{SP}}\otimes \overline{\S^{3\mathrm{SP}}}$, and we may then do a similarity transformation by an appropriate $4 \times 36$-dimensional matrix to obtain the $S$-matrix $\widetilde \S $ of $\cB_\times^{(3)}$, giving
\bea
\widetilde \S = {1\over 1 + \phi^2}\left( \begin{matrix} 
1 & \phi & \phi & \phi^2 \\ 
\phi & -1 & \phi^2 & -\phi \\
\phi & \phi^2 & -1 & - \phi \\ 
\phi^2 & - \phi & - \phi & 1
\end{matrix} \right) ~,\hspace{0.5 in} \phi = {1\over 2}(1 + \sqrt{5})~.
\eea
This we recognize as the $S$-matrix of $\mathrm{Fib} \times \mathrm{Fib}^\mathrm{op}$, i.e.
\bea
\widetilde \S =  \S^\mathrm{Fib} \otimes \S^\mathrm{Fib}~, \hspace{0.5 in} \S^\mathrm{Fib} ={1\over \sqrt{1 + \phi^2}} \left(\begin{matrix} 1 & \phi \\ \phi & - 1\end{matrix}  \right) ~,
\eea
where the opposite superscript on the second Fibonacci factor is identified via  inspection of the $T$-matrix. 

Using the techniques described in  e.g. Appendix B.4 of \cite{Albert:2025umy}, this then allows us to conclude the following:
\begin{itemize}
\item Gauging the binary algebra $\cA^{(3)}_\mathrm{diag} := 1  \boxtimes \bar 1  + W  \boxtimes \overline W$ in $\cB_\times^{(3)}$ gives us the diagonal Monster CFT, where $W$ and $\overline W$ generate the two Fibonacci symmetries;

\item The diagonal Monster theory has a dual $\mathrm{Fib} \boxtimes\mathrm{Fib}^\mathrm{op}$ symmetry, and gauging of the diagonal binary algebra $\widehat \cA_\mathrm{diag}^{(3)}$ returns us to $\cB_\times^{(3)}$.\footnote{Note that this Fibonacci symmetry of the Monster theory has been previously identified in \cite{Fosbinder-Elkins:2024hff,Moller:2024xtt}. } 

\end{itemize}
In a sense, the theory $\cB^{(3)}_\times$ appearing here is the $c=24$ analog of $(\mathfrak{g}_2)_1 \times (\mathfrak{f}_4)_1$ at $c=8$. 

To summarize, what we have found is that\footnote{We also have the more obvious relationship 
\bea
\cM = (\mathrm{3SP} \times \cF_{24})/ \cA_\mathrm{full}^{(3)}~, 
\eea
with 
\bea
\cA_\mathrm{full}^{(3)} = \bigoplus_{\alpha = 0,1} \bigoplus_{\m \in \ZZ_3} L_{\alpha, \mu} \boxtimes \overline{L_{\alpha, \mu}}~, 
\eea
 which follows directly from the bilinear relationship in (\ref{eq:bilinear2}), and which is the analog of the first equation in (\ref{eq:A1A2eqs}). Here
$L_{\alpha, \mu}$ denotes the simple objects of the $\cB(3) = \mathfrak{su}(2)_3/\mathfrak{u}(1)$ chiral category corresponding to the primary $(2 \alpha, 2 \mu)$ (see Section \ref{sec:genericp} for more details), and  $\overline{L_{\alpha, \mu}}$ denotes the matching object in the dual category of $\cB(3)^\vee$. 
} 
\bea
\label{eq:Z3finalresult}
\boxed{\,\,\cM=\cB_\times^{(3)} / \cA^{(3)}_\mathrm{diag}~.\,\,}
\eea
Fermionizing both sides of this, we may also obtain\footnote{We also have the relation 
\bea
{\,\,\mathrm{Paraferm}_{\ZZ_{3A}}\left(\cM / \widehat \cA_\mathrm{diag}^{(3)} \right)  = \cP(3) \times\cP(3)^\vee ~, \,\,}
\eea
where $\widehat \cA_\mathrm{diag}^{(3)}$ is the dual algebra to $\cA^{(3)}_\mathrm{diag}$. } 
\bea
\boxed{\,\,\mathrm{Paraferm}_{\ZZ_{3A}}\left(\cM \right)  = (\cP(3) \times\cP(3)^\vee) /  \cA_\mathrm{diag}^{f,(3)} ~\,\,}
\eea
for an appropriate non-invertible $ \cA_\mathrm{diag}^{f,(3)} $, which is the image of $\cA^{(3)}_\mathrm{diag}$ under the bosonization interface. This algebra is essentially the same as the original one---indeed, since $\mathrm{Fib} \times \mathrm{Fib}^\mathrm{op}$ acts only on the $\ZZ_3$-neutral sector, it survives under parafermionization, as does the algebra. This also implies that, in addition to the modular action reorganizing paraspin sectors, the theory $\cP(3)$ has modular data for $\mathrm{Fib}$, and as a result there is no holomorphic paraspin-CFT realization of it that could be used to write generalizations of (\ref{eq:LinShaoclaim})  and (\ref{eq:MonsterVOAstatement}). This is why we have focused on relations involving diagonal CFTs.

\section{Extension to general $\ZZ_{p A}$}
\label{sec:genericp}

We next generalize the discussion to parafermionization with respect to $\ZZ_{pA}$, for generic odd primes $p$. 
To begin, note that the bosonization $\cB(p) = \mathfrak{su}(2)_p/\mathfrak{u}(1)$ of the $\ZZ_p$-parafermion theory $\cP(p)$ has primaries 
\bea
(\ell, m) ~, \hspace{0.2 in} 0 \leq \ell \leq p~, \hspace{0.2 in} m \in \ZZ_{2p}~, \hspace{0.2 in} \ell + m = 0 \,\,\,\mathrm{mod}\,\,2~,
\eea
subject to identifications 
\bea
(\ell, m) \sim (p - \ell, m+ p)~.
\eea
For odd $p$, a complete set of representations is given by $(2\alpha, 2\mu)$ for $\alpha = 0,\dots, {p - 1 \over 2}$ and $\mu \in  \ZZ_p$. 
The index $\alpha$ labels the $(p+1)/2$ simple-current orbits, while $\mu$ labels the position in the $\ZZ_p$-orbit. 
The characters are given explicitly in Appendix \ref{app:characters}, and we collect each element of an orbit into a vector, 
\bea
\boldsymbol{\chi}_\alpha^{\cB(p)} := (\chi^{\cB(p)}_{(2\alpha,0)}\,,\, \dots\,,\, \chi^{\cB(p)}_{(2\alpha,2(p-1))})~.
\eea

Now assume that the Monster partition function may be decomposed as follows, 
\bea
\label{eq:generalMonsterdecomp}
J(\tau) = \sum_{\alpha=0}^ {(p - 1 )/ 2} \boldsymbol{\chi}_\alpha^{\cB(p)}\cdot  \boldsymbol{\chi}_\alpha^\mathrm{\cB(p)^\vee}~,
\eea
where $\boldsymbol{\chi}_\alpha^{\cB(p)^\vee}$ are a putative set of dual characters; the existence of such a decomposition is our main assumption.
Then defining the tilded characters 
\bea
\widetilde \chi^{(p)}_{\alpha,\beta} : =  \boldsymbol{\chi}_\alpha^{\cB(p)}\cdot  \boldsymbol{\chi}_\beta^{\cB(p)^\vee}~,
\eea
we may rewrite the decomposition as 
\bea
\label{eq:mostgeneralMonster}
|J(\tau)|^2 =  \Big|\sum_{\alpha=0}^{(p-1 )/ 2} \widetilde \chi^{(p)}_{\alpha,\alpha} \Big|^2 ~.
\eea
On the other hand, we will now show that the bosonization $\cB^{(p)}_\times:= \mathrm{Bos}(\cP(p) \times \cP(p)^\vee)$ is given by 
\bea
\label{eq:mainusefulformula}
Z^{\cB^{(p)}_\times} =  \sum_{\alpha,\beta = 0}^{(p-1 )/ 2} | \widetilde \chi^{(p)}_{\alpha,\beta} |^2~, 
\eea
which implies that $\cB^{(p)}_\times$ and $\cM$ can be related by discrete gauging. 

The result (\ref{eq:mainusefulformula}) may be proven as follows. 
Consider more generally an arbitrary pair of bosonic theories $\cB_1$ and $\cB_2$ with $\ZZ_N$ symmetry. 
Call the respective parafermionizations $\cP_1$ and $\cP_2$, which have partition functions
\bea
\label{eq:individualparaferms}
Z^{\cP_i}_{s_1, s_2} = {1\over N} \sum_{a_1,a_2} \omega^{(s_1 + a_1)(s_2 + a_2)} Z^{\cB_i}_{a_1, a_2}~, \hspace{0.5 in} i=1,2~.
\eea
Next consider the bosonization of the product theory $\cP_1 \times \cP_2$. This gives a new bosonic theory $\cB^{(N)}_\times := \mathrm{Bos} (\cP_1 \times \cP_2)$ with partition function 
\bea
Z^{\cB^{(N)}_\times} = {1\over N} \sum_{s_1,s_2} \omega^{-s_1 s_2} Z^{\cP_1}_{s_1, s_2} \, Z^{\cP_2}_{s_1, s_2} ~. 
\eea
Plugging in (\ref{eq:individualparaferms}) and summing over $s_1$ gives a delta function setting $s_2 = - b - d$, and we obtain
\bea
\label{eq:Bdashdef}
Z^{\cB^{(N)}_\times} = {1\over N^2} \sum_{a,b,c,d} \omega^{-ad - bc} Z^{\cB_1}_{a,b} \, Z^{\cB_2}_{c,d}~. 
\eea

Next, we split each bosonic theory $\cB_i$ into sectors, each of which are orbits under the $\ZZ_N$ simple current, and we collect the characters of each entry in the orbit into a vector $\boldsymbol{\chi}_\alpha^{\cB_i}$, 
where $\alpha$ takes values up to the total number of orbits. We assume that this number is the same for theories $\cB_1$ and $\cB_2$. 
Once we have this decomposition, we may write 
\bea
\label{eq:decompositionBi}
Z^{\cB_i}_{a,b} = \sum_{\alpha} \sum_{\mu=0}^{N-1} \omega^{\mu b} (\boldsymbol{\chi}_\alpha^{\cB_i} )_{\mu+a}  \overline{(\boldsymbol{\chi}_\alpha^{\cB_i} )}_{\mu-a} ~, \hspace{0.4 in} i =1,2~.
\eea
Then plugging this into the expression (\ref{eq:Bdashdef}) for $\cB^{(N)}_\times$, we obtain 
\bea
Z^{\cB^{(N)}_\times} = {1\over N^2}  \sum_{a,b,c,d} \sum_{\alpha, \beta} \sum_{\mu,\nu=0}^{N-1} \omega^{-ad - bc} \omega^{\mu b + \nu d} (\boldsymbol{\chi}_{\alpha}^{\cB_1} )_{\mu+a}\overline{(\boldsymbol{\chi}_{\alpha}^{\cB_1} )}_{\mu-a}(\boldsymbol{\chi}_{\beta}^{\cB_2} )_{\nu+c}\overline{(\boldsymbol{\chi}_{\beta}^{\cB_2} )}_{\nu-c}~.
\eea
Doing the sums over $b$ and $d$ produces a pair of delta functions setting $c=\mu $ and $a= \nu$, giving
\bea
Z^{\cB^{(N)}_\times} = \sum_{\alpha, \beta} \sum_{\mu,\nu=0}^{N-1}(\boldsymbol{\chi}_{\alpha}^{\cB_1} )_{\mu + \nu}(\boldsymbol{\chi}_{\beta}^{\cB_2} )_{\mu + \nu}\overline{(\boldsymbol{\chi}_{\alpha}^{\cB_1} )}_{\mu-\nu}\overline{(\boldsymbol{\chi}_{\beta}^{\cB_2} )}_{\nu-\mu}~.
\eea
At this stage we assume that $N$ is odd, in which case 2 is invertible mod $N$, and we may change variables via $\mu\rightarrow \bar 2( \mu + \nu)$ and $\nu \rightarrow \bar 2 (\mu - \nu)$. This gives
\bea
Z^{\cB^{(N)}_\times} = \sum_{\alpha, \beta} \sum_{\mu,\nu=0}^{N-1}(\boldsymbol{\chi}_{\alpha}^{\cB_1} )_{\mu}(\boldsymbol{\chi}_{\beta}^{\cB_2} )_{\mu}\overline{(\boldsymbol{\chi}_{\alpha}^{\cB_1} )}_{\nu}\overline{(\boldsymbol{\chi}_{\beta}^{\cB_2} )}_{-\nu}~. 
\eea
Finally, defining 
\bea
\widetilde \chi^{(N)}_{\alpha, \beta} := \sum_{\mu} (\boldsymbol{\chi}_{\alpha}^{\cB_1} )_\mu (\boldsymbol{\chi}_{\beta}^{\cB_2} )_\mu~, \hspace{0.5 in}\widetilde \chi^{(N)\,C}_{\alpha, \beta} := \sum_{\mu} (\boldsymbol{\chi}_{\alpha}^{\cB_1} )_\mu (\boldsymbol{\chi}_{\beta}^{\cB_2} )_{-\mu}~,
\eea
we see that the bosonization of $\cP_1 \times \cP_2$ is given by 
\bea
\label{eq:producttheorybos}
Z^{\cB^{(N)}_\times} = \sum_{\alpha,\beta}\, \widetilde \chi^{(N)}_{\alpha,\beta} \,\overline{\widetilde \chi^{(N)\, C}_{\alpha,\beta}}~. 
\eea

Thus far the discussion has been quite general. Specializing to the case of  $\cB_1$ being the bosonization of $\cP(p)$ and $\cB_2$ being the bosonization of $\cP(p)^\vee$, we reproduce the result in (\ref{eq:mainusefulformula}), upon noting that the theories in question are charge conjugation invariant. 

Having established (\ref{eq:mainusefulformula}), we now identify precisely which discrete gauging is involved in going between $\cM$ and $\cB_\times^{(p)}$. To do so, we first identify the symmetries of $\cB^{(p)}_\times$. In the basis of primaries $(2\alpha,2\mu)$ given before, the parafermion $S$-matrix is given by 
\bea
\S_{(2\alpha,2\mu), (2\beta,2\nu)} = {2 \over \sqrt{p(p+2)}}\, \omega^{2 \mu\nu}\, \mathrm{sin}\left({(2 \alpha + 1) (2 \beta + 1) \pi \over p + 2} \right) ~,
\eea
which may be split into two factors, 
\bea
\S_{(2\alpha,2\mu), (2\beta,2\nu)} = \S^{(0)}_{\mu,\nu} \times \S^{\mathfrak{so}(3)_p}_{\alpha, \beta}~,
\eea
where 
\bea
\label{eq:SO3pSdef}
\S^{(0)}_{\mu,\nu}:= {1\over \sqrt{p}}\, \omega^{2 \mu \nu}~, \hspace{0.4 in} \S^{\mathfrak{so}(3)_p}_{\alpha, \beta} := {2 \over \sqrt{p+2}}\, \mathrm{sin}\left({(2 \alpha + 1) (2 \beta+ 1) \pi \over p + 2} \right) ~,
\eea
both of which are unitary (since 2 is invertible mod $p$). Note that $\S^{\mathfrak{so}(3)_p}$ is precisely the $S$-matrix of the integer-spin subcategory of $\mathfrak{su}(2)_p$.\footnote{Note that it is $\sqrt{2}$ times the relevant part of the $\mathfrak{su}(2)_p$ $S$-matrix \cite{francesco2012conformal}, with the factor of $\sqrt{2}$ included since the integer sub-category has half the total quantum dimension as the full category. } 

The product theory $\cP(p) \times \cP^\vee(p)$  has $S$-matrix $\S \otimes \overline \S$, and applying this to the tilded characters gives 
\bea
\widetilde \chi^{(p)}_{\alpha, \beta} &\rightarrow& \sum_{\alpha' ,\mu',\beta', \nu'}\sum_\mu \S_{(2 \alpha, 2 \mu), (2 \alpha', 2\mu')} (\boldsymbol{\chi}^{\cB(p)}_{\alpha'})_{\mu'} \overline{\S}_{(2\beta,2\mu),(2 \beta',2\nu')} (\boldsymbol{\chi}^{\cB(p)^\vee}_{\beta'})_{\nu'}
\no\\
&=&  \sum_{\alpha', \mu',\beta',\nu'}  \sum_\mu \S^{(0)}_{\mu,\mu'}\overline{\S^{(0)}_{\mu,\nu'}} \S^{\mathfrak{so}(3)_p}_{\alpha,\alpha'}\S^{\mathfrak{so}(3)_p}_{\beta,\beta'}(\boldsymbol{\chi}^{\cB(p)}_{\alpha'})_{\mu'}  (\boldsymbol{\chi}^{\cB(p)^\vee}_{\beta'})_{\nu'}
\no\\
&=& \sum_{\alpha', \beta'}    \S^{\mathfrak{so}(3)_p}_{\alpha,\alpha'}\S^{\mathfrak{so}(3)_p}_{\beta,\beta'} \sum_{\mu'}(\boldsymbol{\chi}^{\cB(p)}_{\alpha'})_{\mu'}  (\boldsymbol{\chi}^{\cB(p)^\vee}_{\beta'})_{\mu'}
\no\\
&=& \sum_{\alpha',\beta'}\S^{\mathfrak{so}(3)_p}_{\alpha,\alpha'}\S^{\mathfrak{so}(3)_p}_{\beta,\beta'} \widetilde \chi^{(p)}_{\alpha',\beta'}~,
\eea
where we have used unitarity of $\S^{(0)}$. Thus the $S$-matrix acting on the tilded characters is given by 
\bea
\widetilde \S : = \S^{\mathfrak{so}(3)_p} \otimes \S^{\mathfrak{so}(3)_p}~. 
\eea
In a similar way, starting from the parafermion $T$-matrix,
\bea
\T_{(2\alpha,2\mu),(2\alpha,2\mu)} = \mathrm{exp}\left[ 2 \pi i \left( {\alpha(\alpha+1) \over  p+2} - {\mu^2 \over p} - {p-1 \over 12 (p+2) }\right)  \right]~,
\eea
one finds that the $T$-matrix acting on the tilded characters is given by 
\bea
\widetilde \T :=  \T^{\mathfrak{so}(3)_p} \otimes \overline{\T^{\mathfrak{so}(3)_p}}~,
\eea
where $\T^{\mathfrak{so}(3)_p}$ is a diagonal matrix with entries
\bea
\label{eq:Tso3def}
(\T^{\mathfrak{so}(3)_p})_{\alpha, \alpha}  = e^{2 \pi i {\alpha (\alpha+1) \over p+2}}~, \hspace{0.4in} \alpha = 0 , \dots, {p - 1 \over 2}~. 
\eea
In conclusion, the theory $\cB^{(p)}_\times$ has Verlinde lines given by,
\bea
\mathrm{Rep}(\mathfrak{so}(3)_p) \boxtimes \mathrm{Rep}(\mathfrak{so}(3)_p)^\mathrm{op} ~.
\eea

Denoting the simple objects of $\mathrm{Rep}(\mathfrak{so}(3)_p) $ by $L_0, L_1, \dots, L_{p-1 \over 2}$, we then define the canonical diagonal algebra 
\bea
\cA_\mathrm{diag}^{(p)} := \bigoplus_{\alpha=0}^{(p- 1)/ 2} L_\alpha \boxtimes \overline{L_\alpha}~. 
\eea
It follows immediately from (\ref{eq:mostgeneralMonster}) that gauging this object brings us to the Monster theory. This gives us the final result,
\bea
\boxed{\,\,\cM = {\cB^{(p)}_\times/ \cA_\mathrm{diag}^{(p)} }~, \hspace{0.5 in}\cB^{(p)}_\times= \mathrm{Bos}(\cP(p) \times \cP(p)^\vee) ~,\,\,}
\eea
or equivalently 
\bea
\boxed{\,\,\mathrm{Paraferm}_{\ZZ_{pA}} (\cM) = {(\cP(p) \times \cP(p)^\vee) / \cA_\mathrm{diag}^{f, (p)} }~,\,\,} 
\eea
subject to the assumption of a decomposition of the form (\ref{eq:generalMonsterdecomp}). 

Note that in general, for modular $\cC$ the product $\cC \boxtimes \cC^\mathrm{op}$ is the Drinfeld center $\cZ(\cC)$, and the canonical diagonal algebra is Lagrangian, so condensing it gives a holomorphic theory. The dual topological symmetry of the gauged theory is again described by $\cC \boxtimes \cC^\mathrm{op}$ \cite{Mueger:2001crc}. 
Applying this to $\cC = \mathrm{Rep}(\mathfrak{so}(3)_p)$, we conclude that $V^\natural$ has $\mathrm{Rep}(\mathfrak{so}(3)_p) $ symmetry for all $p$ such that the decomposition (\ref{eq:generalMonsterdecomp}) exists.\footnote{In fact, there is an even larger symmetry that comes from the bosonic decomposition 
\bea
\cM = (\cB(p) \times \cB(p)^\vee) /\cA_\mathrm{full}^{(p)}~, \hspace{0.4 in}\cA_\mathrm{full}^{(p)} = \bigoplus_{\alpha = 0}^{(p-1)/2} \bigoplus_{\m \in \ZZ_p} L_{(2\alpha, 2\mu)} \boxtimes \overline{L_{(2\alpha, 2\mu)}}~, 
\eea
though we will not consider the defect McKay-Thompson series for this larger symmetry.}

\section{Defect McKay-Thompson series}
\label{sec:defectMTseries}

Above, we have argued that $V^\natural$ has symmetry $\mathrm{Rep}(\mathfrak{so}(3)_p)$, as long as we have a decomposition of the form (\ref{eq:generalMonsterdecomp}). We now use this symmetry to define new defect McKay-Thompson series, and then compute their invariance subgroups. 

To begin, note that $\mathrm{Rep}(\mathfrak{so}(3)_p)$ for odd $p$ is a rank-${p+1 \over 2}$ symmetry with lines $L_\alpha$ for $\alpha = 0, \dots, {p - 1 \over 2}$ and fusion ring 
\bea
L_\alpha \times L_\beta = \bigoplus_{\gamma = |\alpha - \beta|}^{\mathrm{min}(\alpha + \beta, p - \alpha - \beta)} L_\gamma~. 
\eea
Thus, for example, for $p=3$ we obtain a rank-two fusion category with ring
\bea
L_1 \times L_1 = L_0 + L_1~,
\eea
which are the Fibonacci fusion rules identified previously in Section \ref{sec:Z3warmup}. Likewise, for the case of $p=5$ we obtain a rank-three fusion category with ring, 
\bea
L_1 \times L_1 = L_0 + L_1 + L_2~, \hspace{0.2 in} L_2 \times L_2 = L_0 + L_1 ~, \hspace{0.2 in} L_1 \times L_2 = L_1 + L_2~. 
\eea
The case of $p=7$ is rank four, and its fusion ring is similarly obtained.

We now compute the defect McKay-Thompson series for non-identity elements of the above ring. This gives, 
\bea
T_{L_\alpha}^{(p)}(\tau) := \mathrm{Tr}_{\mathcal{H}}\left( \hat L_\alpha\, q^{L_0 - {c\over 24}}\right) = \sum_{\beta=0}^{p-1\over 2} \,{\S_{\alpha, \beta}^{\mathfrak{so}(3)_p} \over \S_{0, \beta}^{\mathfrak{so}(3)_p} }  \, \widetilde \chi^{(p)}_{\beta, \beta}(\tau)~, \hspace{0.4in} \alpha = 1,\dots, {p-1 \over 2}~.
\eea
Using (\ref{eq:SO3pSdef}), this may be written more explicitly as 
\bea
T_{L_\alpha}^{(p)}(\tau)= \sum_{\beta=0}^{p-1\over 2} {\sin\left({(2 \alpha + 1) (2 \beta + 1) \pi \over p+2 } \right) \over \sin \left({(2 \beta + 1) \pi \over p + 2} \right) } \, \widetilde \chi^{(p)}_{\beta,\beta}(\tau)~,
\eea
where the explicit expressions for the tilded characters $ \widetilde \chi^{(p)}_{\beta,\beta}(\tau)$ can be found in Appendix \ref{app:characters} for $p=3,5,7$.

It now straightforward to obtain e.g.
\bea
T^{(3)}_{L_1}(\tau)&=&\half (1+ \sqrt{5}) \left( \frac{1}{q} +59045\, q + 6092960\, q^2 + 241395800\, q^3+\dots \right)
\no\\
&\vphantom{.}& \hspace{0.2 in}
\half (1- \sqrt{5})\left( 137839\, q + 15400800\, q^2 + 622904170\, q^3+\dots \right)~,
\eea
or equivalently 
\bea
T^{(3)}_{L_1}(\tau)&=&\frac{1}{2} \left( \frac{1}{q} +196884\,q +21493760\,q^2 +864299970\,q^3 +\dots \right)
\no\\
&\vphantom{.}& \hspace{0.2 in}
+\frac{\sqrt{5}}{2} \left( \frac{1}{q} -78794\,q -9307840\,q^2 -381508370\,q^3 +\dots \right)~,
\eea 
for the Fibonacci-like defect at $p=3$. 
This particular defect McKay-Thompson series was previously identified in \cite{Fosbinder-Elkins:2024hff}, where it was shown that the subgroup under which the McKay-Thompson series was invariant was 5$D_0$ in the notation of \cite{cummins2003congruence,onlinetable}, which is equivalent to $\Gamma_1(5)$. 

Similarly for $p=5$, we compute,
\bea
\label{eq:peq5MTseries}
T^{(5)}_{L_1}(\tau)&=&{1\over 2}\csc { \pi \over 14} \left( q^{-1}+36410\,q +3702350\, q^2 +143775457\, q^3 +\dots \right)
\no\\
&\vphantom{.}& \hspace{0.2 in}- \half  \csc {3 \pi \over 14}\left( 86944 q + 10141668 q^2 + 420987784 q^3 +\dots \right)
\no\\
&\vphantom{.}& \hspace{0.2 in}
+ \half \csc{5 \pi \over 14}  \left( 73530 q + 7649742 q^2 + 299536729 q^3 +\dots \right)~,
\no\\
\no\\
T^{(5)}_{L_2}(\tau)&=&2 \sin {5 \pi \over 14} \left(q^{-1} +36410\,q +3702350\, q^2 +143775457\, q^3 +\dots \right)
\no\\
&\vphantom{.}& \hspace{0.2 in}+ 2 \sin { \pi \over 14}\left( 86944 q + 10141668 q^2 + 420987784 q^3 +\dots \right)
\no\\
&\vphantom{.}& \hspace{0.2 in}
-2\sin {3 \pi \over 14}  \left( 73530 q + 7649742 q^2 + 299536729 q^3 +\dots \right)~,
\eea
for the two non-trivial defects of $\mathrm{Rep}(\mathfrak{so}(3)_5)$,
while for $p=7$ we obtain
\bea
\label{eq:peq7MTseries}
T^{(7)}_{L_1}(\tau)&=&{\sqrt{3}\over 2} \csc {\pi \over 9} \left(q^{-1} + 15913 q + 1443200 q^2 + 53104893 q^3 +\dots \right)
\no\\
&\vphantom{.}& \hspace{0.2 in}
-{\sqrt{3}\over 2} \sec{\pi \over 18} \left( 69226 q + 8331976 q^2 + 349080252 q^3 +\dots \right)
\no\\
&\vphantom{.}& \hspace{0.2 in}+ {\sqrt{3}\over 2} \csc {2 \pi \over 9}\left( 46378 q + 4556888 q^2 + 174036624 q^3 +\dots \right)~,
\no\\
T^{(7)}_{L_2}(\tau)&=&{1\over 2} \csc {\pi \over 18} \left(q^{-1} + 15913 q + 1443200 q^2 + 53104893 q^3 +\dots \right)
\no\\
&\vphantom{.}& \hspace{0.2 in}
- \left( 65367 q + 7161696 q^2 + 288078201 q^3 + \dots \right)
\no\\
&\vphantom{.}& \hspace{0.2 in}
+\sec{\pi \over 18} \sin{2 \pi \over 9} \left( 69226 q + 8331976 q^2 + 349080252 q^3 +\dots \right)
\no\\
&\vphantom{.}& \hspace{0.2 in}- \half \sec { \pi \over 9}\left( 46378 q + 4556888 q^2 + 174036624 q^3 +\dots \right)~,
\no\\
T^{(7)}_{L_3}(\tau)&=&2 \cos {\pi \over 9} \left(q^{-1} + 15913 q + 1443200 q^2 + 53104893 q^3 +\dots \right)
\no\\
&\vphantom{.}& \hspace{0.2 in}
+ \left( 65367 q + 7161696 q^2 + 288078201 q^3 + \dots \right)
\no\\
&\vphantom{.}& \hspace{0.2 in}
-2\sin{\pi \over 18}  \left( 69226 q + 8331976 q^2 + 349080252 q^3 +\dots \right)
\no\\
&\vphantom{.}& \hspace{0.2 in}- \cos {\pi \over 18} \csc {2 \pi \over 9}\left( 46378 q + 4556888 q^2 + 174036624 q^3 +\dots \right)~.
\eea

Let us now try to identify the subgroups leaving the $p=5,7$ defect McKay-Thompson series invariant. We may first try to do so using the same technique as in \cite{Fosbinder-Elkins:2024hff}. Namely, requiring that 
\bea
T_{L_\alpha}^{(p)}(\gamma \tau) = T_{L_\alpha}^{(p)}( \tau) ~, \hspace{0.5 in} \gamma \in \mathrm{PSL}(2, \ZZ)~,
\eea
is equivalent to finding the subset of $\gamma$ for which 
\bea
\label{eq:DLorbits}
\rho_{(p)}(\gamma)^\dagger\, D_{L_\alpha}\, \rho_{(p)}(\gamma) = D_{L_\alpha} ~, 
\eea
where $D_{L_\alpha}$ is the diagonal matrix with entries
\bea
(D_{L_\alpha})_{\beta\beta} := {\S_{\alpha, \beta}^{\mathfrak{so}(3)_p} \over \S_{0,\beta}^{\mathfrak{so}(3)_p} }~,\hspace{0.5 in} \alpha,\beta = 0 , \dots {p - 1 \over 2}~,
\eea
and $\rho_{(p)}(\gamma)$ is the projective representation 
\bea
\rho_{(p)} \,\,: \,\, \mathrm{PSL}(2,\ZZ) \rightarrow \mathrm{GL}\left({p+1 \over 2}, \CC\right)
\eea
generated by 
\bea
\rho_{(p)}(S)= \S^{\mathfrak{so}(3)_p} ~, \hspace{0.5 in} \rho_{(p)}(T) = \T^{\mathfrak{so}(3)_p} ~. 
\eea
The strategy of \cite{Fosbinder-Elkins:2024hff} is then to identify all elements in the $\mathrm{SL}(2,\ZZ)$ orbit of $D_{L_\alpha}$ under (\ref{eq:DLorbits}), to represent the actions of $\rho_{(p)}(S)$ and $\rho_{(p)}(T)$ as permutations on these elements, and then to use this cycle structure to read off the invariance subgroup. 

For example, focusing on $p=5$, we find a total of 24 elements in the orbit of  $D_{L_1}$, which can be written as follows,
\bea
D_{0,1} &=& \mathrm{diag}\left( c_1, -c_2, c_3 \right) ~, \hspace{0.2 in} D_{0,2} =  \mathrm{diag}\left( -c_2, c_3, c_1 \right) ~, \hspace{0.2 in} D_{0,3} =  \mathrm{diag}\left( c_3, c_1, -c_2 \right) ~,
\\
D_{1,x} &=& \left(\begin{matrix} 0 & \zeta^x & 0 \\ \zeta^{-x} & 0 & \zeta^{-3x} \\ 0 & \zeta^{3x} & 1\end{matrix} \right)~, 
\hspace{0.1 in}
D_{2,x} = \left(\begin{matrix} 1 &0 & - \zeta^{-x} \\ 0 & 0 & -\zeta^{2x} \\ -\zeta^x & -\zeta^{-2 x} & 0\end{matrix} \right)~, 
\hspace{0.1 in}
D_{3,x} = \left(\begin{matrix} 0 &-\zeta^{-2 x} &  \zeta^{-3x} \\ -\zeta^{2x} & 1 & 0 \\ \zeta^{3x} & 0& 0\end{matrix} \right)~, 
\no
\eea
where  $c_n = 2 \mathrm{cos}{n \pi \over 7}$ and $\zeta = e^{2 \pi i /7}$. The labeling is chosen such that $D_{L_1} = D_{0,1}$ and
\bea
(\S^{\mathfrak{so}(3)_5})^\dagger D_{x,y} \S^{\mathfrak{so}(3)_5} = D_{y,-x}~, \hspace{0.5 in}(T^{\mathfrak{so}(3)_5})^\dagger D_{x,y} \T^{\mathfrak{so}(3)_5} = D_{x,y+x}~.
\eea
In particular, we see that under $\T^{\mathfrak{so}(3)_5}$ we have three fixed elements (namely $D_{0,1} , D_{0,2},$ and $D_{0,3}$) and three 7-cycles. Thus the action of $\T^{\mathfrak{so}(3)_5}$ is represented by the permutation $1^3 7^3$, while there are no fixed points of $\S^{\mathfrak{so}(3)_5}$ and $(\S\T)^{\mathfrak{so}(3)_5}$. Comparing this to the table in \cite{onlinetable}, we conclude that the invariance group is $\Gamma_1(7)$. Note that this is a genus-zero subgroup of  $\mathrm{SL}(2,\ZZ)$. The same result holds for $D_{L_2}$. 

For the case of $p=7$, a similar exercise shows that the invariance group is $\Gamma_1(9)$ for $D_{L_1}, D_{L_2},$ and $D_{L_3}$.
This suggests a general pattern: namely that the McKay-Thompson series for elements of $\mathrm{Rep}(\mathfrak{so}(3)_p)$ have invariance subgroup $\Gamma_1(p+2)$, where we recall that\footnote{By slight abuse of notation, we write $\Gamma_1(N)$ for the image
$\overline{\Gamma}_1(N)$ of the usual congruence subgroup $\Gamma_1(N)\subset \mathrm{SL}(2, \ZZ)$ in $\mathrm{PSL}(2, \ZZ)$.  In the cases considered here this is harmless, since $N$ is odd, so $-\mathds{1} \notin \Gamma_1(N)$, and the projection $\Gamma_1(N)\to \overline{\Gamma}_1(N)$ is injective. 
}
\bea
\Gamma_1 (N) = \left\{ \begin{pmatrix} a & b \\ c & d \end{pmatrix} \,\, \Big | \,\, a, d = 1 \,\,\,\mathrm{mod}\,\, N,~~ c = 0 \,\,\, \mathrm{mod}\,\, N\right\} ~.
\eea
Indeed, this may be proven as follows. 

Denote the stabilizer of the line $L_\alpha$ by 
\bea
\Gamma_\alpha^{(p)} := \left\{\gamma \in \mathrm{PSL}(2, \ZZ) \hspace{0.1 in} | \hspace{0.1 in}T_{L_\alpha}^{(p)}(\gamma \tau) = T_{L_\alpha}^{(p)}( \tau)  \right\} ~.
\eea
We first show that $\Gamma_1(p+2) \subseteq \Gamma_\alpha^{(p)}$ for any $\alpha = 0,\dots, {p-1 \over 2}$. 
To do so, consider any $\gamma = \left( \begin{smallmatrix} a & b \\ c & d \end{smallmatrix} \right) \in \Gamma_1(p+2)$. 
Modulo $p+2$, we have that
\bea
\gamma = \begin{pmatrix} 1 & b \\ 0 & 1 \end{pmatrix} = T^b~, 
\eea
which implies that 
\bea
\gamma \,T^{-b} \in \Gamma(p+2)~. 
\eea
On the other hand, because  $\rho_{(p)}(T) = \T^{\mathfrak{so}(3)_p}$ satisfies $(\rho_{(p)}(T) )^{p+2}=\mathds{1}$,  it follows by the congruence property of modular tensor categories \cite{Ng:2010win,Ng:2012ty} that the projective kernel $\mathrm{ker} \rho_{(p)}$ contains $\Gamma(p+2)$.  This means that $\Gamma(p+2)$ acts via a phase in the projective representation, and we have 
\bea
\rho_{(p)}(\gamma) = \lambda_\gamma (\T^{\mathfrak{so}(3)_p} )^{b}~. 
\eea
But since $D_{L_\alpha}$ and $ \T^{\mathfrak{so}(3)_p}$ are both diagonal and hence commute, we immediately obtain
\bea
\rho_{(p)}(\gamma)^\dagger D_{L_\alpha}  \rho_{(p)}(\gamma) = (\T^{\mathfrak{so}(3)_p})^{-b} D_{L_\alpha} (\T^{\mathfrak{so}(3)_p})^b = D_{L_\alpha}~,
\eea
so that $\Gamma_1(p+2) \subseteq \Gamma_\alpha^{(p)}$ as desired. Note that this argument was rather general and did not use any  properties of $\mathrm{Rep}({\mathfrak{so}}(3)_p)$ besides the order of the $T$-matrix. 

We next show that for non-trivial $\alpha = 1,\dots, {p-1 \over 2}$, we also have the reverse inclusion $ \Gamma_\alpha^{(p)} \subseteq \Gamma_1(p+2)$. The proof of this is significantly more involved. To show it, note that any $\gamma \in  \Gamma_\alpha^{(p)}$ must satisfy
\bea
\rho_{(p)}(\gamma)\, D_{L_\alpha} = D_{L_\alpha}\,\rho_{(p)}(\gamma)~,
\eea
or in components 
\bea
\left(  {\S_{\alpha, \beta}^{\mathfrak{so}(3)_p} \over \S_{0,\beta}^{\mathfrak{so}(3)_p} }  -  {\S_{\alpha, \delta}^{\mathfrak{so}(3)_p} \over \S_{0,\delta}^{\mathfrak{so}(3)_p} } \right) (\rho_{(p)}(\gamma))_{\beta, \delta}= 0~.
\eea
In particular, let us focus on the first column, with $\delta = 0$. In this case, the term in parenthesis is non-zero for any $\beta \neq 0$,\footnote{This follows by noting that the fusion coefficients can be written as matrices $(N_\alpha)_{\beta \gamma} = ( \S^\dagger D_\alpha \S)_{\beta\gamma}$, which means that the entries of $D_\alpha$ are the eigenvalues of $N_\alpha$. Then applying the Perron-Frobenius theorem to $N_\alpha$, which is a non-negative irreducible matrix,  there is a unique largest eigenvalue, and hence a unique largest entry of $D_\alpha$,  which is the quantum dimension $(D_\alpha)_{00}$ of $L_\alpha$. } and hence we must have 
\bea
\label{eq:mainrhoequation}
 (\rho_{(p)}(\gamma))_{\beta, 0}= 0~, \hspace{0.2 in}\forall \beta \neq 0~.
\eea
We will now show that this implies $\gamma \in \Gamma_1(p+2)$. Of course, when $c=0$, then $\pm \gamma$ is simply a power of $\T^{\mathfrak{so}(3)_p}$, and hence $ (\rho_{(p)}(\gamma))$ is a diagonal matrix satisfying (\ref{eq:mainrhoequation}). Now consider $c \neq 0$. 
Our first goal is to show that when $c \neq 0 $ mod $p+2$, then $ (\rho_{(p)}(\gamma))_{\beta, 0}$ is non-vanishing for some $\beta$, and thus that $ \Gamma_\alpha^{(p)} \subseteq \Gamma_0(p+2)$.

First consider the case of $\mathrm{gcd}(c,p+2) = 1$. In this case $c$ is invertible mod $p+2$, and we may decompose 
\bea
\label{eq:gammadecomp}
\gamma = \left(\begin{matrix} 1 & a \bar c \\ 0  &1 \end{matrix} \right)  \left(\begin{matrix} 0 & -1 \\ 1  &0 \end{matrix} \right)  \left(\begin{matrix} c & 0 \\ 0  &\bar c \end{matrix} \right) \left(\begin{matrix} 1 & d \bar c \\ 0  &1 \end{matrix} \right) \,\,\,\mathrm{mod}\,\,p+2~.
\eea
The first and last components are powers of $T$, and hence act diagonally in the representation. They may thus be ignored. 
On the other hand, as discussed in Appendix \ref{app:modular}, the diagonal matrix  $\mathrm{diag}(c, \bar c)$ acts as a permutation, changing the column label via $(2\beta+1) \rightarrow \bar c(2 \beta+1)$. 
Since the $S$-matrix takes the usual form
\bea
(\rho_{(p)}(S))_{\alpha, \beta} = {2 \over \sqrt{p+2}} \sin \left( { (2 \alpha + 1) (2 \beta + 1) \pi \over p+ 2} \right)~, 
\eea
we conclude that, up to an overall factor, we have 
\bea
(\rho_{(p)}(\gamma))_{\beta, 0} &\propto&  \sin \left( { (2 \beta + 1) \bar c\, \pi\over p+ 2} \right)~. 
\eea
But since $\bar c$ is invertible mod $p+2$ and  $(2 \beta + 1) \neq 0$ mod $p+2$, we have $(2 \beta + 1) \bar c \neq 0$ mod $p+2$, and hence we conclude that the entire first column is non-zero, violating (\ref{eq:mainrhoequation}). 

Next consider the case of $c \neq 0 $ mod $p+2$ and $\mathrm{gcd}(c,p+2) >1$. In this case we may no longer use the decomposition in (\ref{eq:gammadecomp}). Instead, note that for $c \neq 0 $ we have the general expression
\bea
\rho_{(p)}(\gamma)_{\beta, 0} \propto K^{(p)}_\gamma(2 \beta + 1,1) - K^{(p)}_\gamma(-2 \beta -1,  1)~,
\eea
where 
\bea
K^{(p)}_\gamma(m,n):= e^{ - 2 \pi i {\bar 4 \over (p+2) c} (a m^2 - 2 m n + d n^2)} \delta_{m, d n}^{(\mathrm{gcd}(c, p+2))}~
\eea
and $\delta_{x,y}^{(N)}$ is a delta-function imposing $x = y$ mod $N$; this result is derived in Appendix \ref{app:modular}. We now show that this is non-vanishing for at least one non-zero $\beta$.  Indeed, first consider any $m$ satisfying
\bea
\label{eq:choiceforbeta}
m = d \,\,\,\mathrm{mod}\,\, \mathrm{gcd}(c, p+2) \hspace{0.3 in}\mathrm{and}\hspace{0.3 in}  m \neq \pm 1\,\,\, \mathrm{mod}\,\, p+2~.
\eea
Such an $m$ always exists. To see this, note that the solutions to the first equation mod $p+2$ are 
\bea
m = d + k\, \mathrm{gcd}(c, p+2)~, \hspace{0.4 in} k = 0 , \dots, {p+2 \over \mathrm{gcd}(c, p+2)} - 1~.
\eea
But since $p+2$ is odd and $\mathrm{gcd}(c, p+2)< p+2$ (because we are assuming $c \neq 0 $ mod $p+2$), we have$ {p+2 \over \mathrm{gcd}(c, p+2)} \geq 3$. Thus there are at least three solutions, and at most only two are $\pm 1$, so we can always choose at least one other solution. Since $p+2$ is odd, for any such $m \in \ZZ_{p+2}$, exactly one of $\{m, -m\}$ is odd, and can be taken to have values in $1, 3, \dots, p$. We then define $\beta$ by writing that odd element as $ 2 \beta + 1$.

For such a $\beta$, the delta function $\delta_{2 \beta + 1, d}^{(\mathrm{gcd}(c,p+2))}$ appearing in the first term clearly gives a non-zero contribution. On the other hand, the second term gives a vanishing contribution, and hence the total difference is non-zero. Indeed, since $-2 \beta -1 = -d \,\,\,\mathrm{mod}\,\, \mathrm{gcd}(c, p+2)$, to get a non-zero contribution from the second term we would have to have $2 d = 0\,\,\,\mathrm{mod}\,\, \mathrm{gcd}(c, p+2)$. But this cannot be satisfied---because $\mathrm{gcd}(c,p+2)$ is odd and greater than 1, we have $\mathrm{gcd}(c,p+2) \nmid 2$, while at the same time $\mathrm{gcd}(c,p+2)\nmid d$ since $\mathrm{gcd}(c, d) = 1$. 
We thus conclude that $ \Gamma_\alpha^{(p)} \subseteq \Gamma_0(p+2)$. 

Our final goal is to show that elements of $\Gamma_\alpha^{(p)} $ must also have $a = d = 1$ mod $p+2$, so that $ \Gamma_\alpha^{(p)} \subseteq  \Gamma_1(p+2)$. Since we already have $c=0$ mod $p+2$, it follows that $d = \bar a$ mod $p+2$, so we may focus on $d$. But for $c=0$ mod $p+2$, the delta functions in $(\rho_{(p)}(\gamma))_{\beta, 0} $ enforce $2 \beta+1 = \pm d$ mod $p+2$, and for this to have non-zero support only on $\beta = 0$, it follows that $d = \pm 1$ mod $p+2$ in $\mathrm{SL}(2, \ZZ)$, i.e. we can take $d=1$ mod $p+2$ projectively. Thus we have that $ \Gamma_\alpha^{(p)} \subseteq  \Gamma_1(p+2)$, and hence that $\Gamma_\alpha^{(p)}=\Gamma_1(p+2)$.

\section*{Acknowledgments}
 YH is supported by Grant-in-Aid for JSPS Fellows No. 25KJ1925. JK is supported by JSPS KAKENHI Grant No. 26K17152 and through the Inamori Frontier Program at Kyushu University. IO is supported by the Forefront Physics and Mathematics Program to Drive Transformation (FoPM), a World-leading Innovative Graduate Study (WINGS) Program, and JSR Fellowship at the University of Tokyo.

\appendix

\section{Fermionization and parafermionization}
\label{app:fermionization}

\subsection{Fermionization}
We begin by reviewing standard fermionization and bosonization.
Although we focus on torus partition functions, the discussion in this
subsection naturally extends to arbitrary closed surfaces. Given a bosonic theory $\mathcal{B}$ with a non-anomalous
$\mathbb{Z}_2$ symmetry, one can construct a fermionic theory whose partition
function depends on the choice of spin structure. We denote the partition
function of $\mathcal{B}$ in the presence of a background gauge field
$A\in H^1(T^2;\mathbb{Z}_2)\cong \ZZ_2\times \ZZ_2$ for this $\mathbb{Z}_2$ symmetry by
$Z^{\mathcal{B}}(A)$.

Let us now recall the Arf invariant, which gives the partition function of the
nontrivial invertible $(1+1)$-dimensional fermionic SPT phase. For a spin
structure $s$ on the torus, specified by
$(s_1,s_2)\in \mathbb{Z}_2\times \mathbb{Z}_2$, we define
\begin{equation}
   \mathrm{Arf}(s):= \mathrm{Arf}(s_1,s_2)=s_1s_2 \mod 2 ~.
\end{equation}
Here $s_i=0,1$ specifies the NS/R boundary condition along the $i$-th cycle of
the torus, respectively.
Then we can also define the Arf invariant for a spin structure shifted by a
$\mathbb{Z}_2$ background field $A=(A_1,A_2)$ as
\begin{equation}
    \mathrm{Arf}(s+A)
    :=
    \mathrm{Arf}(s_1+A_1,s_2+A_2)
    =
    (s_1+A_1)(s_2+A_2)
    \mod 2 ~.
\end{equation}
This is well-defined because $s+A$ is again a spin structure.\footnote{
In general, the set of spin structures on $\Sigma_g$ is a torsor over
$H^1(\Sigma_g;\mathbb{Z}_2)$. Thus, there is no canonical choice of origin, 
but after choosing a reference spin structure $s$, any other spin
structure can be written as $s+A$ with
$A\in H^1(\Sigma_g;\mathbb{Z}_2)$.
}

Then fermionization can be written as
\begin{equation}
\label{eq:fermionizationdef}
    Z^{\mathcal{F}}(s)
    =
    \frac{1}{2}
    \sum_{a\in H^1(T^2;\mathbb{Z}_2)}
    (-1)^{\mathrm{Arf}(s+a)}
    Z^{\mathcal{B}}(a) ~,
\end{equation}
which we also denote by $Z^\cF_{s_1, s_2}$, while bosonization is given by the inverse transformation,
\begin{equation}
    Z^{\mathcal{B}}(A)
    =
    \frac{1}{2}
    \sum_{s}
    (-1)^{\mathrm{Arf}(s+A)}
    Z^{\mathcal{F}}(s) ~,
\end{equation}
where the sum is over all spin structures on $T^2$.\footnote{
Here we assume that the gravitational anomaly vanishes mod $16$. Under this assumption, gauging $(-1)^F$ is equivalent to
summing over spin structures, and either description may be regarded as
bosonization. See~\cite{BoyleSmith:2024qgx} for more details on this subtlety.
}

One can also obtain a fermionic theory in another way. Namely, we first gauge
the original $\mathbb{Z}_2$ symmetry,
\begin{equation}
    Z^{\mathcal{B}_{\mathrm{gauged}}}(\widehat A)
    =
    \frac{1}{2}
    \sum_{a\in H^1(T^2;\mathbb{Z}_2)}
    (-1)^{\int \widehat A \cup a}
    Z^{\mathcal{B}}(a) ~,
\end{equation}
where $\widehat A$ is the background gauge field for the dual $\widehat{\ZZ}_2$ symmetry,
and then fermionize this gauged theory with respect to the dual symmetry,
\begin{equation}
    Z^{\widetilde{\mathcal{F}}}(s)
    =
    \frac{1}{2}
    \sum_{\widehat a\in H^1(T^2;\mathbb{Z}_2)}
    (-1)^{\mathrm{Arf}(s+\widehat a)}
    Z^{\mathcal{B}_{\mathrm{gauged}}}(\widehat a) ~.
\end{equation}
A direct computation shows that these two ways of fermionizing the theory differ
by stacking with the Arf theory,
\begin{equation}
    \label{eq:fermionstack}
    Z^{\widetilde{\mathcal{F}}}(s)
    =
    (-1)^{\mathrm{Arf}(s)}
    Z^{\mathcal{F}}(s) ~.
\end{equation}

\subsection{Parafermionization}
By analogy with the above discussion, one can consider a $\mathbb{Z}_k$ version
of fermionization, namely parafermionization~\cite{Yao:2020dqx,Thorngren:2021yso}. 
We note that this operation is defined only at the level of torus partition functions; see the comments below for more details on this point. 

Given a bosonic theory $\mathcal{B}$ with a non-anomalous
$\mathbb{Z}_k$ symmetry, one can construct a collection of parafermionic torus partition functions labeled by the choice of paraspin structure.
Taking the trivial paraspin structure on the torus to be $s_1=s_2=0$, we define
a $\mathbb{Z}_k$-valued Arf-like quadratic phase by
\begin{equation}
    \mathrm{Arf}_k(s):=s_1s_2\mod k
\end{equation}
where $s_1,s_2\in\mathbb{Z}_k$.\footnote{Note that this quantity depends on the choice of basis for the cycles of the torus and does not define the partition function of an invertible paraspin TQFT.}
Then one can define parafermionized torus partition function as
\begin{equation}
\label{eq:parafermionization}
    Z^{\cP}(s):=\frac{1}{k}\sum_{a\in H^1(T^2;\ZZ_k)}\omega^{\mathrm{Arf}_k(s+a)}Z^{\cB}(a)~,
\end{equation}
where $\omega:=e^{2\pi i/k}$. We also denote this by $Z^\cP_{s_1, s_2}$. 
Similarly to the previous subsection, bosonization can be defined as
\begin{equation}
    Z^{\cB}(A):=\frac{1}{k}\sum_{s}\omega^{-\mathrm{Arf}_k(s+A)}Z^{\cP}(s)~.
\end{equation}
By the same reasoning as for fermionization, we may also parafermionize $\cB_\mathrm{gauged}$ to get  $\widetilde \cP$ with
\begin{equation}
    Z^{\widetilde\cP}(s)=\omega^{\mathrm{Arf}_k(s)}Z^{\cP}(s')~,
\end{equation}
where now $s'=(-s_1,s_2)$ for $s=(s_1,s_2)$. In contrast to the fermionic relation (\ref{eq:fermionstack}), this should not be interpreted as stacking with an invertible paraspin TQFT; it is a relation among torus partition functions.

\paragraph{Comment on paraspin structures:} There are several subtleties in defining parafermionization:
\begin{itemize}
    \item A $\mathbb{Z}_k$ paraspin structure is naturally a two-dimensional
    notion. More precisely, it can be viewed as a lift of the $SO(2)$ frame
    bundle to its $k$-fold cover. Such a lift exists only when the Euler
    characteristic of the surface is divisible by $k$; in particular, it always
    exists on the torus.
    \item It is not clear, at least from the torus partition function alone,
    whether the resulting parafermionization defines a local QFT on more
    general backgrounds.
\end{itemize}

Beginning with the first point, recall that a spin structure on an
oriented manifold $M$ is a lift of the $SO(n)$ frame bundle to a
$\mathrm{Spin}(n)$ bundle. Equivalently, in a cochain description, after choosing
a representative of the second Stiefel--Whitney class
$w_2(TM)\in H^2(M;\mathbb{Z}_2)$, a spin structure may be described as a
$\mathbb{Z}_2$-valued $1$-cochain $\eta\in C^1(M;\mathbb{Z}_2)$ satisfying
\begin{equation}
    \delta \eta = w_2(TM)~.
\end{equation}
Thus a spin structure exists if and only if $w_2(TM)=0$ in cohomology, and the
set of spin structures, when nonempty, is a torsor over
$H^1(M;\mathbb{Z}_2)$.

For paraspin structures, it is therefore natural to ask whether there is an
analogous characteristic class whose trivialization specifies a
$\mathbb{Z}_k$ paraspin structure. However, the relevant obstruction is not
usually phrased as a $\mathbb{Z}_k$-valued Stiefel--Whitney class. Rather, for an
oriented surface, a $\mathbb{Z}_k$ paraspin structure can be regarded as a lift
of the $SO(2)$ frame bundle through the $k$-fold cover
$\mathrm{Spin}_k(2)\to SO(2)$ such that $\mathrm{Spin}_k(2)/\ZZ_k=SO(2)$. The obstruction is the Euler class, or equivalently the first
Chern class of the tangent bundle, reduced modulo $k$.

This low-dimensional interpretation also has a physical counterpart. In
$(2+1)$ dimensions, massive particles are classified by representations of the
little group $SO(2)$. Since this group admits arbitrary $k$-fold covers, it is
natural that parafermionic degrees of freedom can appear as massive
excitations, namely Abelian anyons, in two-dimensional topological phases. By
contrast, in higher dimensions the relevant rotation groups are $SO(n)$ with
$n\geq 3$, for which the spin double cover is the natural universal cover.
This suggests that $\mathbb{Z}_k$ paraspin structures are not expected to have
an equally natural higher-dimensional analogue. See~\cite{Radicevic:2018okd,geiges2010generalisedspinstructures2dimensional,Randal_Williams_2013,Runkel:2018feb} for more details.

Regarding the second point, 
according to e.g.~\cite{Gaiotto:2015zta}, fermionic signs and the dependence on a
spin structure can be implemented by a Grassmann integral. This gives a local
state-sum-like manipulation: the additional variables are assigned locally, and
invariance under changes of triangulation is governed by local algebraic
identities. In this sense, fermionization is not merely a prescription for
reorganizing torus partition functions; it has a locality-preserving
implementation.

For $\mathbb{Z}_k$ parafermionization with $k>2$, the analogous statement is
less clear. Existing constructions are often formulated in terms of generalized
Jordan--Wigner transformations, attachment of a parafermionic phase, or
relations among torus partition functions. While these constructions are useful
and suggestive, they do not by themselves provide a direct analogue of the fermionization, nor do they immediately establish a local
state-sum or path-integral formulation on paraspin backgrounds.

\subsection{Additional comments}
\label{app:newappendix}

In Section \ref{sec:clarification}, we claimed that if $V$ is a VOA admitting an order-2 simple current extension $FV$, and if $\cV$ and $\cF \cV$ are the respective diagonal (spin-)CFTs, one may show that $\cF \cV$ is obtained from $\cV$ via fermionization in the sense of (\ref{eq:fermionizationdef}). 
To see this, say that our simple current extension is by a $\ZZ_2$-current $J$. Label the characters of $\cV$ as $\chi^\cV_{\alpha, \mu}$, where $\alpha$ labels $\ZZ_2$-orbits, and $\mu \in \ZZ_{\ell_\alpha}$, where $\ell_\alpha = 2$ for non-trivial orbits and $\ell_\alpha = 1 $ for fixed points. Define the charges 
\bea
\eps_\alpha : = e^{2 \pi i (h_{\alpha, \mu+1} - h_J - h_{\alpha, \mu})} \in \{\pm 1 \}~,
\eea
where $h_J \in \half + \ZZ$ so that in particular $\eps_\alpha = -1$ for fixed points. 
We then  compute the following twisted-twined partition functions, 
\bea
Z^{\cV}_{0,0} &=& \sum_{\alpha, \mu} |\chi^\cV_{\alpha, \mu}|^2 ~, \hspace{0.5 in} Z^{\cV}_{0,1} \,\,=\,\, \sum_{\alpha, \mu} \eps_\alpha |\chi^\cV_{\alpha, \mu}|^2 ~, 
\no\\
 Z^{\cV}_{1,0} &=& \sum_{\alpha, \mu } \chi^\cV_{\alpha, \mu} \overline {\chi^\cV_{\alpha, \mu + 1}}~, \hspace{0.24 in}  Z^{\cV}_{1,1} \,\,=\,\, -  \sum_{\alpha, \mu } \eps_\alpha \chi^\cV_{\alpha, \mu} \overline {\chi^\cV_{\alpha, \mu + 1}}~,
\eea
or altogether 
\bea
\label{eq:generalZcV}
Z^{\cV}_{a,b} = \sum_{\alpha, \mu} (-1)^{ab} \eps_\alpha^b\, \chi^\cV_{\alpha, \mu}\, \overline{\chi^\cV_{\alpha, \mu + a}}~.
\eea
Next, fermionize this partition function via  (\ref{eq:fermionizationdef}).
Doing so gives,
\bea
Z^{\cF\cV}_{s_1, s_2} &=& \half (-1)^{s_1 s_2} \sum_{a}\sum_{\alpha, \mu} (-1)^{a s_2} \left(\sum_b \eps_\alpha^b (-1)^{b s_1} \right) \chi^\cV_{\alpha, \mu}\, \overline{\chi^\cV_{\alpha, \mu + a}}
\no\\
&=& (-1)^{s_1 s_2} \sum_{\alpha : \,\, \eps_\alpha = (-1)^{s_1}}  \sum_\mu \left[ |\chi_{\alpha, \mu}|^2 + (-1)^{s_2}   \chi^\cV_{\alpha, \mu} \,\overline {\chi^\cV_{\alpha, \mu + 1}} \right] ~.
\eea
Splitting the non-trivial orbits from the fixed points, we obtain,
\bea
\label{eq:generalcFcV}
Z^{\cF\cV}_{s_1, s_2} = (-1)^{s_1 s_2} \sum_{\substack{\alpha :\,\, \ell_\alpha =2 \\ \eps_\alpha = (-1)^{s_1}}} | \chi_{\alpha, 0} + (-1)^{s_2} \chi_{\alpha, 1}|^2 + \delta_{s_1, 1} (1+(-1)^{s_2}) \sum_{\alpha :\,\, \ell_\alpha = 1} |\chi_{\alpha, 0}|^2~,
\eea
which is precisely what one would expect for the diagonal spin-CFT constructed from the sVOA $FV = V \oplus J$.\footnote{For example, in the Ising case we have one orbit with $\ell_\alpha = 2$ and $\eps_\alpha = 1$, containing $\chi^\mathrm{Ising}_0$ and $\chi^\mathrm{Ising}_{1/2}$, as well as a single fixed point given by  $\chi^\mathrm{Ising}_{1/16}$. Then (\ref{eq:generalcFcV}) reproduces the expressions in (\ref{eq:diagonalMW}). } 

On the other hand, analogous statements do not hold for parafermionization. To show this, it suffices to focus on cases with no $J$-fixed points. Say that $J$ has $h_J \in -{1 \over p} + \ZZ$, where $p$ is an odd prime. The analog of (\ref{eq:generalZcV}) is then\footnote{This notation is slightly different than that used in (\ref{eq:decompositionBi}); the two are related by relabeling $a \rightarrow - 2 a$ and $b \rightarrow - \bar 2b$ and modifying our definition of the characters $\chi_{\alpha, \mu + \bar 2 q_\alpha}\rightarrow (\boldsymbol{\chi}_{\alpha})_\mu$.   } 
\bea
Z^{\cV}_{a,b} = \sum_{\alpha, \mu}\omega^{-(2\mu + a) b}  \eps_\alpha^b\, \chi^{\cV}_{\alpha, \mu}\, \overline{\chi^{\cV}_{\alpha, \mu+a}}~,
\eea
where $\eps_\alpha$ is now an appropriate $\ZZ_p$ charge, which we may write as $\eps_\alpha = \omega^{q_\alpha}$. Then applying the parafermionization defined in (\ref{eq:parafermionization}) and repeating the same steps as above, we obtain 
\bea
Z^{\cP\cV}_{s_1,s_2} = \omega^{ s_1 s_2} \sum_\alpha \sum_a \omega^{a s_2} \chi^\cV_{\alpha, \overline{2}( s_1 + q_\alpha)} \overline{  \chi^\cV_{\alpha,\overline{2}(s_1 + q_\alpha)+a} }~,
\eea
where $\overline {2}$ is the modular inverse of $2$ mod $p$, which exists since $p$ is odd.  Importantly, for each $\alpha, s_1$, the sum above takes a character with $\mu =  \overline{2}( s_1 + q_\alpha)$ on one side and pairs it with all $J^a$-translates on the other side. Thus the result is not the square of an extended character,  and hence is not the naive diagonal paraspin-CFT constructed from some tentative $\ZZ_p$-graded chiral extension $PV$. 

\section{Explicit results}
\label{app:characters}

In this appendix, we collect various explicit results for the cases of $p=3,5,7$.
Note that in general, the bosonization $\cB(p)$ of the $\ZZ_p$-parafermion theory $\cP(p)$ has central charge \cite{Fateev:1985mm,Thorngren:2021yso}
\bea
c = {2 (p-1) \over p+2}~
\eea
and primaries labelled by 
\bea
(\ell, m) ~, \hspace{0.2 in} 0 \leq \ell \leq p~, \hspace{0.2 in} m \in \ZZ_{2p}~, \hspace{0.2 in} \ell + m = 0 \,\,\,\mathrm{mod}\,\,2~,
\eea
subject to identifications 
\bea
(\ell, m) \sim (p - \ell, m+ p)~,
\eea
with corresponding conformal dimensions
\bea
h_{\ell,m} = {\ell (\ell + 2) \over 4 (p+2)} - {m^2 \over 4 p}~. 
\eea
From this we obtain the $T$-matrix, 
\bea
\T_{(\ell,m),(\ell,m)} = \mathrm{exp}\left[ 2 \pi i \left( {\ell(\ell+2) \over  4(p+2)} - {m^2 \over 4 p} - {p-1 \over 12 (p+2) }\right)  \right]~.
\eea
On the other hand, the $S$-matrix is given by 
\bea
\label{eq:appSmat}
\S_{(\ell,m), (k,n)} = {2 \over \sqrt{p(p+2)}}\, \omega^{{m n  / 2}}\, \mathrm{sin}\left({(\ell + 1) (k + 1) \pi \over p + 2} \right) ~.
\eea
The characters admit the following explicit expression, 
\bea
\chi^{\cB(p)}_{\ell, m}(\tau) &=& {q^{- {c-2 \over 24} + h_{\ell,m}} \over \eta(\tau)^2} \sum_{r,s} (-1)^{r+s} q^{\half s (s+1) + \half r(r+1)+ r s (p+1)}
\no\\
&\vphantom{.}& \hspace{0.5 in} \times \left( q^{\half s (\ell - m) + \half r (\ell + m)}  - q^{p+1- \ell + \half s (2 p + 2 - \ell + m) + \half r (2 p + 2 - \ell - m)} \right) ~.
\eea

\subsection{$p=3$}
\label{app:p3}

In the case of $p=3$, the primaries may be labelled by 
\bea
(\ell, m) \in \left\{ (1,1), (2,0), (2,2), (3,-1), (3,1), (3,3)\right\}~,
\eea
with the corresponding characters given by 
\bea
\chi^{\cB(3)}_{1,1}(\tau) &=&  q^{1/30}(1 + q + 2\, q^2+ 3\, q^3 + \dots )~,
\no\\
\chi^{\cB(3)}_{2,0}(\tau) &=&  q^{11/30}(1 + 2\,q + 2\, q^2+ 4\, q^3 + \dots )~,
\no\\
\chi^{\cB(3)}_{3,\pm1}(\tau) &=&  q^{19/30}(1 + q + 2\, q^2+ 2\, q^3 + \dots )~,
\no\\
\chi^{\cB(3)}_{3,3}(\tau) &=&  q^{- 1/30}(1 + q^2+ 2\, q^3 + 3\, q^4 \dots )~,
\eea
together with $\chi^{\cB(3)}_{2,2} = \chi^{\cB(3)}_{1,1}$.
Note that in Section \ref{sec:Z3warmup}, we used an alternative notation, labeling each character by their corresponding conformal weight, so that 
\bea
\chi_0^\mathrm{3SP} = \chi^{\cB(3)}_{3,3}~, \hspace{0.2 in} \chi_{{2\over 5}}^\mathrm{3SP} = \chi^{\cB(3)}_{2,0}~,\hspace{0.2 in} \chi_{2\over 3}^\mathrm{3SP} = \chi^{\cB(3)}_{3,\pm1} ~, \hspace{0.2 in} \chi_{1\over 15}^\mathrm{3SP}  = \chi^{\cB(3)}_{1,1}=\chi^{\cB(3)}_{2,2}~,
\eea
but this is purely notational. 

By applying the Hecke operator $T_{29}$ to these,\footnote{Hecke operators acting on vector-valued modular forms were defined in \cite{Harvey:2018rdc,Harvey:2019qzs}.} we get a series of dual characters 
\bea
\chi^{\cB(3)^\vee}_{1,1}(\tau) &=&  q^{29\over 30}  (64584 + 6789393\, q + 261202536\, q^2 +5863550310\,q^3 +  \dots)~,
\no\\
\chi^{\cB(3)^\vee}_{2,0}(\tau) &=&  q^{{19\over 30}}  (8671 + 1675504\, q + 83293626\, q^2 + 2175548448\, q^3 + \dots)~,
\no\\
\chi^{\cB(3)^\vee}_{3,\pm1}(\tau) &=&  q^{11\over 30} (783 + 306936\, q + 19648602\, q^2 + 589705488\,q^3+  \dots )~,
\no\\
\chi^{\cB(3)^\vee}_{3,3}(\tau) &=&  q^{-{29\over 30}} (1 + 57478\, q^2 + 5477520\, q^3 + 201424111\, q^4 + \dots)~,
\eea
and as shown in \cite{Bae:2020pvv} these satisfy a bilinear relation of the form 
\bea
J(\tau) = \sum_{\ell, m} \chi^{\cB(3)}_{\ell, m}\, \chi^{\cB(3)^\vee}_{\ell, m}~.
\eea

To rewrite this in a form more consistent with our general discussion in Section \ref{sec:genericp}, 
we may define the following character vectors
\bea
 \boldsymbol{\chi}^{\cB(3)}_0 &=& (\chi^{\cB(3)}_{0,0},\, \chi^{\cB(3)}_{0,2},\,\chi^{\cB(3)}_{0,4}) \,\,=\,\,  (\chi^{\cB(3)}_{3,3},\,\chi^{\cB(3)}_{3,-1},\,\chi^{\cB(3)}_{3,1})~,
 \no\\
 \boldsymbol{\chi}^{\cB(3)}_1 &=& (\chi^{\cB(3)}_{2,0},\,\chi^{\cB(3)}_{2,2},\,\chi^{\cB(3)}_{2,4})\,\,=\,\, (\chi^{\cB(3)}_{2,0},\,\chi^{\cB(3)}_{2,2},\,\chi^{\cB(3)}_{1,1})~,
\eea
with the analogs for $\cB(3)^\vee$. We then define the tilded characters 
\bea
\widetilde \chi^{(3)}_{\alpha, \beta} =  \boldsymbol{\chi}^{\cB(3)}_\alpha \cdot  \boldsymbol{\chi}^{\cB(3)^\vee}_\beta~,
\eea
which are given explicitly by 
\bea
\widetilde \chi^{(3)}_{0,0}(\tau) &=& q^{-1} + 59045\, q + 6092960\, q^2 + 241395800\, q^3 +\dots ~,
\no\\
\widetilde \chi^{(3)}_{0,1}(\tau) &=&q^{3/5} ( 8671  + 1804672\, q+ 97010251\, q^2 + 2713483488\, q^3 + \dots )~,
\no\\
\widetilde \chi^{(3)}_{1,0}(\tau) &=& q^{-3/5}(1 + 1568\, q + 672918\, q^2 + 45506688\, q^3 \dots)~,
\no\\
\widetilde \chi^{(3)}_{1,1}(\tau) &=&137839\, q + 15400800\, q^2 + 622904170\, q^3 + \dots~.
\eea
In particular, we see that we may rewrite the bilinear relation as 
\bea
J(\tau) = \widetilde \chi^{(3)}_{0,0}(\tau) + \widetilde \chi^{(3)}_{1,1}(\tau)~,
\eea
which is the result given in (\ref{eq:MonstertildechiZ3}).

\subsection{$p=5$}

We next consider $p=5$. In \cite{Bae:2020pvv}, it was shown that there exists a bilinear relation of the form 
\bea
J(\tau) = \sum_\alpha \chi_\alpha (\tau) \chi^{VHN^\natural}_\alpha(\tau)~, 
\eea 
where $\chi_\alpha (\tau)$ were certain linear combinations of the characters of $\cB(5) \times \cB(5)$, and the characters  $\chi^{VHN^\natural}_\alpha(\tau)$ were found to take the following form,
\bea
\chi^{VHN^\natural}_0(\tau) &=& q^{-19/21} (1 + 18316\, q^2 + 1360096\, q^3 + 42393826\, q^4+\dots) ~,
\no\\
\chi^{VHN^\natural}_1(\tau) &=&q^{5/21} (133 + 65968\, q + 4172476\, q^2 + 119360584\, q^3+\dots)~,
\no\\
\chi^{VHN^\natural}_2(\tau) &=&q^{5/21} (133 + 65968\, q + 4172476\, q^2 + 119360584\, q^3+\dots)~,
\no\\
\chi^{VHN^\natural}_3(\tau) &=&q^{17/21} (8778 + 1003408\, q + 37866696\, q^2 +\dots)~,
\no\\
\chi^{VHN^\natural}_4(\tau) &=&q^{17/21} (8778 + 1003408\, q + 37866696\, q^2 + \dots)~,
\no\\
\chi^{VHN^\natural}_5(\tau) &=& q^{8/21} (760 + 231705\, q + 12595936\, q^2 + 333082540\, q^3+\dots)~,
\no\\
\chi^{VHN^\natural}_6(\tau) &=&q^{20/21} (35112 + 3184818\, q + 108781232\, q^2 +\dots)~,
\no\\
\chi^{VHN^\natural}_7(\tau) &=&q^{20/21} (35112 + 3184818\, q + 108781232\, q^2 +\dots)~,
\no\\
\chi^{VHN^\natural}_8(\tau) &=&q^{11/21} (3344 + 680504\, q + 32364068\, q^2 + 795272512\, q^3+\dots)~.
\eea
These are expected to be the characters of a well-defined VOA $VHN^\natural$ with symmetry given by the Harada-Norton group $HN$. 

We would now like to repackage this as a bilinear relation of the form 
\bea
\label{eq:peq5bilinear}
J(\tau) = \sum_{\alpha = 0}^2 \boldsymbol{\chi}^{\cB(5)}_\alpha \cdot \boldsymbol{\chi}^{\cB(5)^\vee}_\alpha~.
\eea
This may be done by simply reorganizing the combinations of characters of $\cB(5)$ and  ${VHN^\natural}$ appearing in  \cite{Bae:2020pvv}, so that concretely\footnote{We use the notation $ \chi^{\cB(5)}_{- 1, - 1} =  \chi^{\cB(5)}_{4,4}$ and $ \chi^{\cB(5)}_{-2,-2} =  \chi^{\cB(5)}_{3,3} $. } 
\bea
\chi^{\cB(5)^\vee}_{1,1} &:=& \chi^{\cB(5)}_{5,3} \chi^{VHN^\natural}_2  + \chi^{\cB(5)}_{4,-2}\chi^{VHN^\natural}_5  + \chi^{\cB(5)}_{3,3}\chi^{VHN^\natural}_7~,
\no\\
\chi^{\cB(5)^\vee}_{2,0} &:=&\chi^{\cB(5)}_{2,0} \chi^{VHN^\natural}_8 +  \chi^{\cB(5)}_{4,0} (\chi^{VHN^\natural}_6 + \chi^{VHN^\natural}_7) ~,
\no\\
\chi^{\cB(5)^\vee}_{2,2} &:=&\chi^{\cB(5)}_{5,1} \chi^{VHN^\natural}_4 +  \chi^{\cB(5)}_{1,1} \chi^{VHN^\natural}_6 + \chi^{\cB(5)}_{3,-1} \chi^{VHN^\natural}_8~,
\no\\
\chi^{\cB(5)^\vee}_{3,\pm 1} &:=&\chi^{\cB(5)}_{5,\mp 3} \chi^{VHN^\natural}_4 +  \chi^{\cB(5)}_{4, \mp 2} \chi^{VHN^\natural}_7 + \chi^{\cB(5)}_{\pm 2, \pm 2} \chi^{VHN^\natural}_8~,
\no\\
\chi^{\cB(5)^\vee}_{4,\pm 2} &:=&\chi^{\cB(5)}_{5,\pm 1} \chi^{VHN^\natural}_1 +  \chi^{\cB(5)}_{\pm 1, \pm 1} \chi^{VHN^\natural}_5 + \chi^{\cB(5)}_{3, \pm 1} \chi^{VHN^\natural}_6~,
\no\\
\chi^{\cB(5)^\vee}_{4,0} &:=&\chi^{\cB(5)}_{4,0} \chi^{VHN^\natural}_5~,
\\
\chi^{\cB(5)^\vee}_{5, \pm 3} &:=&\chi^{\cB(5)}_{5,\pm 1} \chi^{VHN^\natural}_0 +  \chi^{\cB(5)}_{\mp 1,  \mp1} \chi^{VHN^\natural}_1 + \chi^{\cB(5)}_{3, \pm 1} \chi^{VHN^\natural}_3~,
\no\\
\chi^{\cB(5)^\vee}_{5, \pm 1} &:=&\chi^{\cB(5)}_{5,\mp 3 } \chi^{VHN^\natural}_0 +  \chi^{\cB(5)}_{4,\mp2} \chi^{VHN^\natural}_2 + \chi^{\cB(5)}_{\mp 2, \mp 2} \chi^{VHN^\natural}_3~,
\no\\
\chi^{\cB(5)^\vee}_{5, 5} &:=&\chi^{\cB(5)}_{5,5 } \chi^{VHN^\natural}_0 +  \chi^{\cB(5)}_{4,0} (\chi^{VHN^\natural}_1 +\chi^{VHN^\natural}_2)+ \chi^{\cB(5)}_{ 2, 0} (\chi^{VHN^\natural}_3+\chi^{VHN^\natural}_4)~,
\no
\eea
as well as 
\bea
\chi^{\cB(5)^\vee}_{4,4} = \chi^{\cB(5)^\vee}_{1,1}~, \hspace{0.5 in} \chi^{\cB(5)^\vee}_{3,3} = \chi^{\cB(5)^\vee}_{2,2}~.
\eea
At the level of $q$-expansions, this gives 
\bea
\chi^{\cB(5)^\vee}_{1,1}(\tau) &=& q^{104/105} (36005 + 3519256\, q + 129371722\, q^2 +\dots)~,
\no\\
\chi^{\cB(5)^\vee}_{2,0}(\tau) &=& q^{16/21}(3344 + 757416\, q + 40245192\, q^2 + 1092644856\, q^3 + \dots)~, 
\no\\
\chi^{\cB(5)^\vee}_{2,2}(\tau) &=& q^{101/105}(38456 + 3915900\, q + 146786912\, q^2 +\dots)~, 
\no\\
\chi^{\cB(5)^\vee}_{3,\pm1}(\tau)&=& q^{59/105}(3344 + 727738\, q + 37321832\, q^2 + 983838676\, q^3+\dots)~,
\no\\
\chi^{\cB(5)^\vee}_{4,\pm 2}(\tau) &=& q^{41/105}(760 + 267710\, q + 16150304\, q^2 + 465675085\, q^3 +\dots)~,
\no\\
\chi^{\cB(5)^\vee}_{4, 0}(\tau) &=& q^{25/21}(760 + 233225\, q + 13062386\, q^2 + 359205792\, q^3 +\dots)~,
\no\\
\chi^{\cB(5)^\vee}_{5,\pm3}(\tau)&=& q^{26/105}(134 + 74880\, q + 5277993\, q^2 +\dots)~,
\no\\
\chi^{\cB(5)^\vee}_{5,\pm1}(\tau)&=& q^{-16/105}(1 + 8912\, q + 1096738\, q^2 + 44579664\, q^3 +\dots)~,
\no\\
\chi^{\cB(5)^\vee}_{5,5}(\tau) &=& q^{-20/21}(1 + 36139\, q^2 + 3534494\, q^3 + 130821726\, q^4+\dots)~. 
\eea
Then defining the character vectors
\bea
\boldsymbol{\chi}_1^{\cB(5)} &:=&(\chi^{\cB(5)}_{0,0},\chi^{\cB(5)}_{0,2},\chi^{\cB(5)}_{0,4},\chi^{\cB(5)}_{0,6},\chi^{\cB(5)}_{0,8}) \,\,=\,\,  (\chi^{\cB(5)}_{5,5},\chi^{\cB(5)}_{5,-3},\chi^{\cB(5)}_{5,-1},\chi^{\cB(5)}_{5,1},\chi^{\cB(5)}_{5,3})~, 
\no\\
\boldsymbol{\chi}_2^{\cB(5)} &:=& (\chi^{\cB(5)}_{2,0},\chi^{\cB(5)}_{2,2},\chi^{\cB(5)}_{2,4},\chi^{\cB(5)}_{2,6},\chi^{\cB(5)}_{2,8}) \,\,=\,\,  (\chi^{\cB(5)}_{2,0},\chi^{\cB(5)}_{2,2},\chi^{\cB(5)}_{3,-1},\chi^{\cB(5)}_{3,1},\chi^{\cB(5)}_{3,3})~, 
\no\\
\boldsymbol{\chi}_3^{\cB(5)} &:=& (\chi^{\cB(5)}_{4,0},\chi^{\cB(5)}_{4,2},\chi^{\cB(5)}_{4,4},\chi^{\cB(5)}_{4,6},\chi^{\cB(5)}_{4,8}) \,\,=\,\, (\chi^{\cB(5)}_{4,0},\chi^{\cB(5)}_{4,2},\chi^{\cB(5)}_{4,4},\chi^{\cB(5)}_{1,1},\chi^{\cB(5)}_{4,-2})~,
\no
\eea
and likewise for $\cB(5)^\vee$, we see that the bilinear relation (\ref{eq:peq5bilinear}) indeed holds.

Next we compute the tilded characters
\bea
\widetilde \chi^{(5)}_{\alpha, \beta} = \boldsymbol{\chi}_\alpha^{\cB(5)} \cdot \boldsymbol{\chi}_\beta^{\cB(5)^\vee} ~,
\eea
whose $q$-expansions are as follows,
\bea
\widetilde \chi^{(5)}_{1,1}(\tau) &=& q^{-1} + 36410\, q + 3702350\, q^2 + 143775457\, q^3+ \dots ~,
\no\\
\widetilde \chi^{(5)}_{1,2}(\tau) &=& q^{5/7}  ( 3344+ 841016\, q + 49619412\, q^{2} +\dots )~,
\no\\
\widetilde \chi^{(5)}_{1,3}(\tau) &=& q^{8/7} ( 2280+ 842175\, q+ 53012736\, q^2+ \dots)  ~,
\no\\
\widetilde \chi^{(5)}_{2,1}(\tau) &=& q^{-5/7} (1 + 272\, q + 203998\, q^2 + \dots) ~,
\no\\
\widetilde \chi^{(5)}_{2,2}(\tau) &=& 86944\, q + 10141668\, q^2 + 420987784\, q^3 + \dots ~,
\no\\
\widetilde \chi^{(5)}_{2,3}(\tau) &=&q^{3/7} ( 1520  + 609710\, q + 40257865\, q^2+\dots)~,
\no\\
\widetilde \chi^{(5)}_{3,1}(\tau) &=& q^{-1/7} ( 3 + 18096\, q + 2397743\, q^2+  \dots )~,
\no\\
\widetilde \chi^{(5)}_{3,2}(\tau) &=& q^{4/7} (  6688  + 1542420\, q + 84862244\, q^2 + \dots ) ~,
\no\\
\widetilde \chi^{(5)}_{3,3}(\tau) &=&73530\, q + 7649742 q^2\, + 299536729\, q^3+ \dots ~.
\eea
The diagonal components are the main ingredients in the defect McKay-Thompson series in (\ref{eq:peq5MTseries}). 

\subsection{$p=7$}

We finally consider the case of $p=7$. In \cite{Bae:2020pvv}, an argument was given for the existence of a VOA with global symmetry given by the Held group $He$, and with characters obtained by application of the Hecke operator $T_{17}$ to $\chi^{\cB(7)}$. This first gives a set of characters $T_{17} \chi^{\cB(7)}$ which satisfy a bilinear relation of the form 
\bea
J(\tau) = \sum_{\alpha, \beta} (T_{17} \chi^{\cB(7)})_\alpha M_{\alpha, \beta}\, \chi^{\cB(7)}_\beta
\eea
for a non-trivial pairing matrix $M$. Since we would like to have a bilinear relation with trivial pairing matrix, we then take the dual character to have entries $\chi^{\cB(7)^\vee}_\beta = \sum_\alpha M_{\alpha, \beta}(T_{17} \chi^{\cB(7)})_\alpha$, which gives explicitly
\bea
\chi^{\cB(7)^\vee}_{1,1}(\tau) &=&  q^{127/126} (22984 + 2144856\, q + 76669456\, q^2 + 1628223858\, q^3+\dots)~,
\no\\
\chi^{\cB(7)^\vee}_{2,0}(\tau) &=&q^{5/6}   (11679 + 1432080\, q + 57855030\, q^2 + 1323345654\, q^3+\dots)~,
\no\\
\chi^{\cB(7)^\vee}_{2,2}(\tau) &=& q^{41/42}  (26112 + 2548623\, q + 93075204\, q^2 + 2002645050\, q^3+\dots)~,
\no\\
\chi^{\cB(7)^\vee}_{3,\pm 1}(\tau) &=&  q^{85/126} (5084 + 836961\, q + 38286006\, q^2 + 941558481\, q^3+\dots)~,
\no\\
\chi^{\cB(7)^\vee}_{3,3}(\tau) &=& q^{121/126}(27234 + 2721020\, q + 100447832\, q^2 + 2175623416\, q^3+\dots)~,
\no\\
\chi^{\cB(7)^\vee}_{4,\pm 2}(\tau) &=&  q^{67/126} (1955 + 448902\,q + 23218498\, q^2 + 612398140\, q^3+\dots)~,
\no\\
\chi^{\cB(7)^\vee}_{4,0}(\tau) &=&  q^{7/18}   (680 + 234226\, q + 13898520\, q^2 + 395054092\, q^3+\dots)~,
\no\\
\chi^{\cB(7)^\vee}_{5,\pm 3}(\tau)&=&  q^{17/42}  (681 + 221646\, q + 12948390\, q^2 + 364895820\, q^3+\dots)~,
\no\\
\chi^{\cB(7)^\vee}_{5,\pm 1}(\tau) &=& q^{5/42}   (51 + 55284\, q + 4454595\, q^2 + 148169790\, q^3+\dots)~,
\no\\
\chi^{\cB(7)^\vee}_{6,\pm 4}(\tau) &=&  q^{37/126} (204 + 97563\, q + 6392374\, q^2 + 191594522\, q^3+\dots)~,
\no\\
\chi^{\cB(7)^\vee}_{6,\pm 2}(\tau) &=& q^{-17/126}(1 + 10404\, q + 1210230\, q^2 + 47772074\, q^3+\dots)~,
\no\\
\chi^{\cB(7)^\vee}_{6,0}(\tau)&=&  q^{13/18}  (4454 + 668848\, q + 29441076\, q^2 + 708076656\, q^3+\dots)~,
\no\\
\chi^{\cB(7)^\vee}_{7,\pm 5}(\tau) &=& q^{25/126} (51 + 32504\, q + 2381394\, q^2 + 75441121\, q^3+\dots)~,
\no\\
\chi^{\cB(7)^\vee}_{7,\pm 3}(\tau) &=& q^{79/126} (1275 + 236980\, q + 11269249\, q^2 + 283574263\, q^3+\dots)~,
\no\\
\chi^{\cB(7)^\vee}_{7,\pm 1}(\tau) &=&  q^{43/126} (153 + 65212\, q + 4054534\, q^2 + 118328160\, q^3+\dots)~,
\no\\
\chi^{\cB(7)^\vee}_{7,7}(\tau) &=& q^{-17/18} (1 + 15810\, q^2 + 1375232\, q^3+47653839\, q^4 \dots)    ~,
\eea
as well as 
\bea
\chi^{\cB(7)^\vee}_{6,6} = \chi^{\cB(7)^\vee}_{1,1}~, \hspace{0.5 in} \chi^{\cB(7)^\vee}_{5,5} = \chi^{\cB(7)^\vee}_{2,2}~, \hspace{0.5 in} \chi^{\cB(7)^\vee}_{4,4} = \chi^{\cB(7)^\vee}_{3,3}~.
\eea
The tilded characters are computed as before, and here we write out only the diagonal components, 
\bea
\widetilde \chi^{(7)}_{1,1}(\tau) &=& q^{-1}+ 15913\, q + 1443200\, q^2 + 53104893\, q^3 +\dots~,
\no\\
\widetilde \chi^{(7)}_{2,2}(\tau) &=& 65367\, q + 7161696\, q^2 + 288078201\, q^3 + \dots ~,
\no\\
\widetilde \chi^{(7)}_{3,3}(\tau) &=& 69226\, q + 8331976\, q^2 + 349080252\, q^3 + \dots~,
\no\\
\widetilde \chi^{(7)}_{4,4}(\tau) &=& 46378\, q + 4556888\, q^2 + 174036624\, q^3 + \dots~,
\eea
which are used for computing the defect McKay-Thompson series in (\ref{eq:peq7MTseries}). 

\section{A finite Weil model for the $\mathrm{Rep}(\mathfrak{so}(3)_p)$ representation }
\label{app:modular}

In order to identify the invariance subgroups of defect McKay-Thompson series associated to $\mathrm{Rep}(\mathfrak{so}(3)_p)$, we require the matrix representation $\rho_{(p)}(\gamma)$ of arbitrary elements $\gamma \in \mathrm{PSL}(2, \ZZ)$. 
One way to obtain this would be direct analysis of the $\mathrm{Rep}(\mathfrak{so}(3)_p)$ characters, but since the closed for expression for these characters is unknown, we instead obtain it in a roundabout way. 
In particular, note that the representation is generated entirely by $\rho_{(p)}(S) = \S^{\mathfrak{so}(3)_p}$ and $\rho_{(p)}(T) = \T^{\mathfrak{so}(3)_p}$, and is $(p+1)/2$-dimensional. If we can identify another system which reproduces these $S$- and $T$-matrices,  the representation of arbitrary $\gamma$ in that system is guaranteed to be equivalent to $\rho_{(p)}(\gamma)$. Below we do this using the Weil representation of $ \mathrm{PSL}(2, \ZZ)$. 

\subsection{The Weil representation }

Given a finite abelian group $G$ and a quadratic form $q: G \rightarrow \mathbb{Q}/\ZZ$, one may define an $N$-dimensional representation of $\mathrm{SL}(2, \ZZ)$ known as the Weil representation in the following way \cite[Section 4]{scheithauer2009weil}, 
\bea
\T^{(G,q)} |m\rangle  = e^{- 2 \pi i q(m)}|m\rangle ~, \hspace{0.5 in} \S^{(G,q)}|m\rangle = {e^{2 \pi i \, \mathrm{sign}(q)/8} \over \sqrt{|G|}}\sum_n e^{2 \pi i B(m, n)} |n\rangle~,
\eea
where $m,n \in \ZZ_N$, $\mathrm{sign}(q)$ is the signature of the quadratic form $q$,\footnote{The signature can be computed modulo 8 via Milgram's formula, 
\bea
e^{2 \pi i\, \mathrm{sign}(q)/8}={1\over \sqrt{|G|}} \sum_m e^{2 \pi i q (m)}~.
\eea
} and $B(m, n)$ is the bilinear form constructed from the quadratic form, namely 
\bea
B(m, n) = q(m + n) - q (m) - q (n)~.
\eea
In this representation the matrix for an arbitrary element $\gamma \in \mathrm{SL}(2, \ZZ)$ may be computed, and is given up to an overall $m, n$-independent phase by \cite[Theorem 1]{stromberg2013weil} (see also \cite{scheithauer2009weil,prasad2009character})
\bea
\rho_{(G,q)}(\gamma)_{m, n} \propto \left\{ \begin{matrix}  \sqrt{|G_c| \over |G|}\, e^{2 \pi i \left(a c\, q(m')  + a\, B(x_c, m') - b d\, q(n) + b\, B(n, m) \right) }&& \exists\, m' \in G\,\,\,\, \mathrm{s.t.} \,\,\,\,m = d n + x_c + c m'
\\
0 && \mathrm{otherwise} \end{matrix}\right.
\no\\
\eea
where $x_c \in G$ is an element of order 2, and $G_c$ is the set of all elements in $G$ with order dividing $c$. 

Let us now consider the specific case of $G= \ZZ_N$ with $N$ odd, and the quadratic form given by 
\bea
q(m) := -{\bar 4 m^2 \over N}~, \hspace{0.4 in} m \in \ZZ_N~,
\eea
where $\bar 4$ satisfying $4 \bar 4 = 1$ mod $N$ exists since $\mathrm{gcd}(4, N) = 1$. The resulting bilinear form and signature are found to be 
\bea
b(m, n) = -{\bar 2 m n \over N}~, \hspace{0.5 in} \mathrm{sign}(q) = \left\{ \begin{matrix} 0 && N = 1\,\,\,\mathrm{mod}\,\,4~, 
\\
6 && N = 3 \,\,\,\mathrm{mod}\,\,4~,
 \end{matrix}\right. 
\eea
and hence the modular $T$- and $S$-matrices take the following form, 
\bea
\T^{(G,q)} |m\rangle  = e^{ 2 \pi i {\bar 4 m^2 \over N}}|m\rangle ~, \hspace{0.5 in} \S^{(G,q)}|m\rangle = {(-i)^{N-1} \over \sqrt{N}}\sum_n e^{-2 \pi i  {\bar 2 m n \over N}} |n\rangle~. 
\eea
Since $N$ is odd, it has no order-2 elements, giving $x_c = 0$, and since 
\bea
G_c =\left \{ x \in G \,\,|\,\, x c = 0 \,\,\, \mathrm{mod}\,\,N \right\}~
\eea
we have $|G_c| = \mathrm{gcd}(c,N)$. Thus we obtain the general expression
\bea
\label{eq:Weilrho1}
\rho_{(G,q)}(\gamma)_{m, n} \propto \left\{ \begin{matrix}  \sqrt{\mathrm{gcd}(c,N)\over N}\, e^{-2 \pi i {\bar 4 \over N } \left(a c (m')^2  - b d n^2 + 2 bn m \right) }&& \exists\, m' \in \ZZ_N\,\,\,\, \mathrm{s.t.} \,\,\,\,m = d n + c m'
\\
0 && \mathrm{otherwise} \end{matrix}\right.~.
\no\\
\eea
When $c$ is non-zero, we may conveniently rewrite this as follows,
\bea
\label{eq:Weilrho}
\rho_{(G,q)}(\gamma)_{m, n} \propto \left\{ \begin{matrix}  \sqrt{\mathrm{gcd}(c,N)\over N}\, e^{-2 \pi i {\bar 4 \over N c} \left(a m ^2 - 2 m n + d n^2 \right) }&&  m = d n \,\,\,\mathrm{mod}\,\,\mathrm{gcd}(c,N)
\\
0 && \mathrm{otherwise} \end{matrix}\right.~,
\eea
where we have used that $ad - b c = 1$ and $m = d n + cm'$. 

\subsection{Connection to $\mathrm{Rep}(\mathfrak{so}(3)_p)$}

The representation discussed so far has been $N$-dimensional. We may now consider the reduction of this representation on the set of anti-symmetrized states $ |m\rangle - |- m \rangle$ for non-trivial $m$. 
This gives an $(N-1)/2$-dimensional representation. Because $N$ is odd, every non-zero sign pair $\{m , - m\}$ has exactly one representative among $1,3, 5, \dots, N-2$, which can be written as $2 \alpha + 1$ for $\alpha = 0, \dots, {N- 3 \over 2}$. We choose this as the label for our anti-symmetrized states, writing them as $| \alpha \rangle \rangle $. 
We then have
\bea
\T^{(G,q)}_\mathrm{odd}|\alpha \rangle \rangle   &=& e^{2 \pi i\,  \bar 4/N}e^{2 \pi i\alpha (\alpha+1) \over N}|\alpha \rangle \rangle  ~, 
\no\\
\S^{(G,q)}_\mathrm{odd}|\alpha\rangle\rangle  &=&- i^{N} {2  \over \sqrt{N}}\sum_\beta \mathrm{sin}\left({(2 \alpha + 1) (2 \beta + 1) \pi \over N} \right)  |\beta\rangle\rangle~.
\eea
When $N=p+2$, we see that these match (up to an overall $\alpha,\beta$-independent proptionality factor) with $\T^{\mathfrak{so}(3)_p}$ and $\S^{\mathfrak{so}(3)_p}$. Consequently, because these generate the full representation, this allows us to read off the representation $\rho_{(p)}(\gamma)$ for a generic $\gamma \in \mathrm{PSL}(2, \ZZ)$ from (\ref{eq:Weilrho1}) and (\ref{eq:Weilrho}), up to a proportionality factor. In particular, for non-zero  $c$ we obtain
\bea
\rho_{(p)}(\gamma)_{\alpha, \beta} \propto K^{(p)}_\gamma(2 \alpha + 1, 2 \beta + 1) - K^{(p)}_\gamma(-2 \alpha -1, 2 \beta + 1)~,
\eea
where 
\bea
K^{(p)}_\gamma(m,n):= e^{ - 2 \pi i {\bar 4 \over (p+2) c} (a m^2 - 2 m n + d n^2)} \delta_{m, d n}^{(\mathrm{gcd}(c, p+2))}~.
\eea
The delta function $\delta_{m, d n}^{(\mathrm{gcd}(c, p+2))}$ above is non-zero when $m = d n $ mod $\mathrm{gcd}(c, p+2)$. In particular, when $c = 0$ mod $p+2$, we require $m = d n $  mod $p+2$. Since $d$ is invertible mod $p+2$ (because $\mathrm{gcd}(c,d) = 1$) this is just a permutation of the labels $m$. 

\bibliographystyle{ytamsalpha}

\baselineskip=.97\baselineskip

\bibliography{ref}

\end{document}